# The FAIREr Guiding Principles: Organizing data and metadata into semantically meaningful types of FAIR Digital Objects to increase their human explorability and cognitive interoperability

Vogt, Lars[1]

[1]*TIB Leibniz Information Centre for Science and Technology, Welfengarten 1B, 30167 Hanover,*

*Germany,* orcid.org/0000-0002-8280-0487

Corresponding Author: lars.m.vogt@googlemail.com

# Abstract

## Background

Ensuring the FAIRness (Findable, Accessible, Interoperable, Reusable) of data and metadata is an important goal in both research and industry. Knowledge graphs and ontologies have been central in achieving this goal, with interoperability of data and metadata receiving much attention. The European Open Science Cloud initiative distinguishes in its Interoperability Framework (EOSC IF) four layers of interoperability: technical, semantic, organizational, and legal.

## Results

This paper argues that the emphasis on machine-actionability has overshadowed the essential need for human-actionability of data and metadata, and provides three examples that describe the lack of human-actionability within knowledge graphs. In response, the paper propagates the incorporation of cognitive interoperability as another vital layer within the EOSC IF and discusses the relation between human explorability of data and metadata and their cognitive interoperability. It suggests adding the Principle of human Explorability, extending FAIR to form the FAIREr Guiding Principles. The subsequent sections present the concept of semantic units, elucidating their important role in enhancing the human explorability and cognitive interoperability of knowledge graphs. Semantic units structure a knowledge graph into identifiable and semantically meaningful subgraphs, each represented with its own resource that constitutes a FAIR Digital Object (FDO) and that instantiates a corresponding FDO class. Various categories of FDOs are distinguished. Each semantic unit can be displayed in a user interface either as a mind-map-like graph or as natural language text.

## Conclusions

Semantic units organize knowledge graphs into levels of representational granularity, distinct granularity trees, and diverse frames of reference. This organization supports the cognitive interoperability of data and metadata and facilitates their human explorability. The development of innovative user interfaces enabled by FDOs that are based on semantic units would empower users to access, navigate, and explore information in knowledge graphs with enhanced efficiency.

# Introduction

Since their initial publication in 2014 (1,2), the **FAIR Guiding Principles** for scientific data management and stewardship have gained significant recognition and are now regarded as essential across different domains of research and industry. The principles serve as a foundation for establishing policies, standards, and practices that ensure the **F**indability, **A**ccessibility, **I**nteroperability, and **R**eusability (FAIR) of scientific data and metadata, benefiting both **machines and humans alike**. Addressing major societal challenges such as biodiversity loss and climate change (3) necessitates the collection, integration, and analysis of large volumes of data from various sources and stakeholders in a truly interdisciplinary approach, which would significantly benefit from the relevant data and metadata being FAIR (4).

The FAIR Guiding Principles provide the basis for the ongoing development of the European Open Science Cloud (EOSC) for shaping a FAIR ecosystem (5,6). To this end, the notion of **FAIR Digital Objects (FDOs)** has garnered attention as a central conceptual foundation within EOSC (7). Each FDO is distinguished by a Globally Unique Persistent and Resolvable Identifier (GUPRI), adheres to common and preferably open file formats, is accompanied by further GUPRIs that provide richly documented metadata adhering to standards and established vocabularies, and includes contextual documentation that, in combination, guarantees an FDO's reliable discovery, citation, and reuse. Some FDOs form containers for particular sets of finer-grained FDOs. Consequently, FDOs can exist at different levels of granularity, allowing for hierarchical structures of nested FDOs.

As atomic entities of the FAIR ecosystem, FDOs take a central role in the **EOSC Interoperability Framework (EOSC IF)** (6) by providing, together with suitable tools and controlled vocabularies such as ontologies, the necessary concepts for achieving interoperability. The overarching goal of this framework, when embedded in a FAIR ecosystem, is to establish a general standard in science and industry. This standard holds the potential to address various ongoing issues in scientific research, including the reproducibility crisis (8), and to enhance the overall trustworthiness of information as enshrined in the TRUST Principles of Transparency, Responsibility, User Focus, Sustainability, and Technology (9).

However, the realization of a FAIR ecosystem hinges on the availability of practical and reliable tools capable of technical implementation. In this context, the strategic utilization of **ontologies** and **knowledge graphs (KGs)**, complemented by a consistent application of **semantic data schemata**, emerges as a promising technological avenue (10–15). Notably, KGs, underpinned by graph-based abstractions, are said to surpass relational or other NoSQL models in several aspects: (i) they provide an intuitive representation of relationships akin to mind-maps, which most users are familiar with; (ii) they facilitate flexible evolution of data schemata, which is particularly valuable for handling incomplete knowledge; (iii) they employ formalisms like ontologies and rules for machine-actionable knowledge representation; (iv) they facilitate graph analytics and machine learning techniques; and (v) they utilize specialized graph query languages supporting not only standard relational operators such as joins, unions, and projections, but also navigational operators for recursively searching entities along arbitrary-length paths (13,16–21). KGs, ontologies, and semantic data schemata have the potential to enhance transparency in data-driven decision-making processes and improve communication across various domains, including research, science, and industry.

While KGs, ontologies, and semantic data schemata hold great promise for enabling data and metadata interoperability and being essential cornerstones in a Going FAIR strategy, they also come with their own technical, conceptual, and societal challenges. For instance, defining the boundaries of the concept of a KG remains somewhat ambiguous (13), considering that KGs encompass various technical and conceptual incarnations, such as labeled property graphs (e.g., Neo4J) and approaches based on the Resource Description Framework (RDF), the use of RDF-stores, and the application of description logics using the Web Ontology Language (OWL). Not all KGs use ontologies (or other controlled vocabularies) and pre-defined semantic data schemata, resulting in **varying degrees of FAIRness and semantic interoperability within a KG and across different KGs**. Moreover, there has been a predominant focus on achieving FAIRness of data and metadata for machines, potentially overlooking the importance of FAIRness for human users and software developers interacting with the KGs.

This paper addresses the human-actionability aspect of FAIRness of data and metadata and discusses in the *Problem statement* section three critical challenges of KGs that arise when focussing predominantly on machine-actionability.

In the *FAIREr Guiding Principles* section, I argue that the **cognitive interoperability** of data and metadata for human users and software developers should be given due consideration within the EOSC IF, and I emphasize the importance of knowledge graph exploration in the context of cognitive interoperability. I argue for the addition of **human explorability of data and metadata** as another basic principle to the FAIR Principles, leading to the proposal of the **FAIREr Guiding Principles** (FAIR + human **E**xplorability **r**aised).

In the *Semantic units as FAIR Digital Objects: a strategy to go FAIREr* section, I provide a brief introduction to the concept of **semantic units** and in the section *How semantic units make FAIR knowledge graphs FAIREr*, I discuss how KGs organized into semantic units and corresponding FDOs meet the criteria on data and metadata specified in the FAIREr Guiding Principles, resulting in **FAIREr KGs**. I also discuss how the organization into semantic units facilitates the development of new UIs that support different exploration strategies that increase the human explorability and cognitive interoperability of data and metadata.

In the *Conclusion* section, I discuss the contributions that semantic units, the FAIREr Guiding Principles, and the addition of cognitive interoperability to the EOSC IF potentially provide for solving the three challenges introduced in the *Problem statement* section and how **semantic units as FDOs** contribute to the bigger picture of Going FAIR, and which **FAIR services** are needed for a **FAIR ecosystem**.

---

**Box 1 | Conventions**

This paper refers to FAIR KGs as machine-actionable semantic graphs used for documenting, organizing, and representing lexical, assertional (e.g., empirical data), contingent, and universal statements, combining instance graphs and class axioms from ontology classes. This distinguishes KGs from ontologies as I understand them, with ontologies primarily containing universal statements (class axioms) and lexical statements. The discussion in this paper revolves around KGs and ontologies within the context of RDF-based triple stores, OWL, and description logics as a formal framework for inferencing. Labeled property graphs can be considered as an alternative to triple stores. These are considered, due to their widespread use, as technologies and logical frameworks for KGs that are supported by a broad community of users and developers and for which accepted standards exist. While acknowledging the existence of alternative technologies and frameworks, such as those supporting n-tuples syntax and more advanced logics like first order logic (22,23), it is important to note that these alternatives lack comprehensive tool support and widespread usage, thus hindering their transformation into well-supported, scalable, and easily usable KG applications.

Throughout this paper, regular underlined text is used to indicate ontology classes, *italicsUnderlined* text when referring to properties (i.e., relations in Neo4j), and ID numbers are included to specify each of them. ID numbers are composed of the ontology prefix followed by a colon and a number (e.g., *isAbout* (IAO:0000136)). In cases where a term is not covered in any ontology, it is denoted by an asterisk (*), for example, the class *<u>metric measurement statement unit</u>*. Instances of classes are indicated using '<u>regular underlined</u>' text, where the label denotes the class label and the ID number corresponds to the class. Furthermore, the term *resource* refers to something uniquely designated, such as a GUPRI, about which you want to say something. It thus stands for something and represents something you want to talk about. In RDF, the *Subject* and the *Predicate* in a triple statement are always resources, while the *Object* can be either a resource or a literal. Resources can encompass properties, instances, or classes, with properties occupying the *Predicate* position in a triple, instances referring to individuals (=particulars), and classes representing universals.

     For the sake of clarity, resources are represented in the text and in figures using human-readable labels instead of their GUPRIs, with the implicit assumption that each resource possesses its own GUPRI.

# Problem statement

## Challenge 1: Machines need more information than humans for data to be actionable, and humans do not want to look at large and complex graphs

Humans possess an innate ability to deal with **contextuality** and **vagueness** of statements using metaphors, metonymies[1], or general figures of thought that are based on conventions and personal experience. As a result, humans often have no problem with **missing information** when communicating with each other because they still understand the intended content. Machines, on the other hand, demand explicit information to process data. This contrast becomes evident in scenarios where concise statements, imbued with contextual nuances, suffice for human communication, but machines require comprehensive and explicit data representations. For instance, someone living in New York has no difficulties understanding the following dialogue (24):

> *A: How did you get to JFK airport?*
> *B: I stopped a cab.*

    Taken literally, however, the dialogue makes no sense because *B* would not have left the spot just by stopping a taxi. An untrained machine would have substantial difficulties identifying the implied meaning of what *B* responded, because important contextual information is missing. A human reader, on the other hand, projects the dialogue back onto a prototypical sequence of events of which stopping the taxi is just the initial of a multitude of steps, and understands what has been said as referring to a well-known general figure of thought. Another example is the following sentence:

> *In Berlin, Germany, on the 10th of March 2021, the 7-days incidence value for COVID-19 was 57.9.*

    Human readers with a background in medicine would have no difficulties understanding this sentence. From the context, they would interpret *"in Berlin, Germany"* to be a somewhat vague

---

[1] A metonymy is the substitution of the name of an attribute or adjunct for that of the thing meant, for example, *"the pen is mightier than the sword"* with *"the pen"* referring to writing and *"the sword"* to warfare and violence.

reference to the human population of the German city of Berlin. A machine would have a hard time carrying out such interpretations without providing additional rules and information.

One consequence of being experts in dealing with contextuality, vagueness, and missing information is that most humans prefer statements to be concise to make communication more time-efficient, with lots of information not being explicitly stated, but implied via context or somewhat vague sentences and by the sender of the information relying on the receiver to possess the relevant background knowledge to correctly decipher the information. In the end, **efficient communication** between people is only possible through this metonymic understanding of language (24).

Besides the need to make all required information explicit for machines, considerations regarding how to best query specific information and how to make semantic data schemata more reusable also frequently influence choices for modelling data and may result in an increase in the complexity of the graph in order to modularize and standardize graph patterns. Therefore, when human-readable statements are being translated into machine-actionable and easy to query representations that follow, for example, the RDF triple syntax paradigm of *Subject-Predicate-Object*, these graph-based representations are often much more complex than human readers like them to be, often containing triples that are not semantically meaningful for a domain expert (e.g., Fig. 1).

We are thus dealing with a dilemma that arises from the **conflict between machine-actionability and human-actionability: the more we push data representations towards machine-actionability, the more complex and thus the less human-actionable they become**. Humans do not want to see large and complex graphs, because these graphs usually contain lots of information that is irrelevant for humans and that, in bulk, reduce the graph's comprehensibility. This dilemma represents an impedance mismatch.

If we want to make the data and metadata of KGs FAIR for both machines *and* humans, we need to make them easier for humans to comprehend. In other words, we must **increase the cognitive interoperability of data and metadata**, for instance by providing new means of exploring and navigating the graph in the form of a **mind-map like graph** (e.g., Fig. 1, bottom), filtering out information that the user is currently not interested in or that is only required for their machine-actionability, and by providing multiple entry-points for exploring the graph, zooming in and out across different levels of representational granularity.

## Observation:

*ObjectX weighs 5 kilograms, with a 95% confidence interval of 4.54 to 5.55*

## Machine-Actionable RDF Graph:

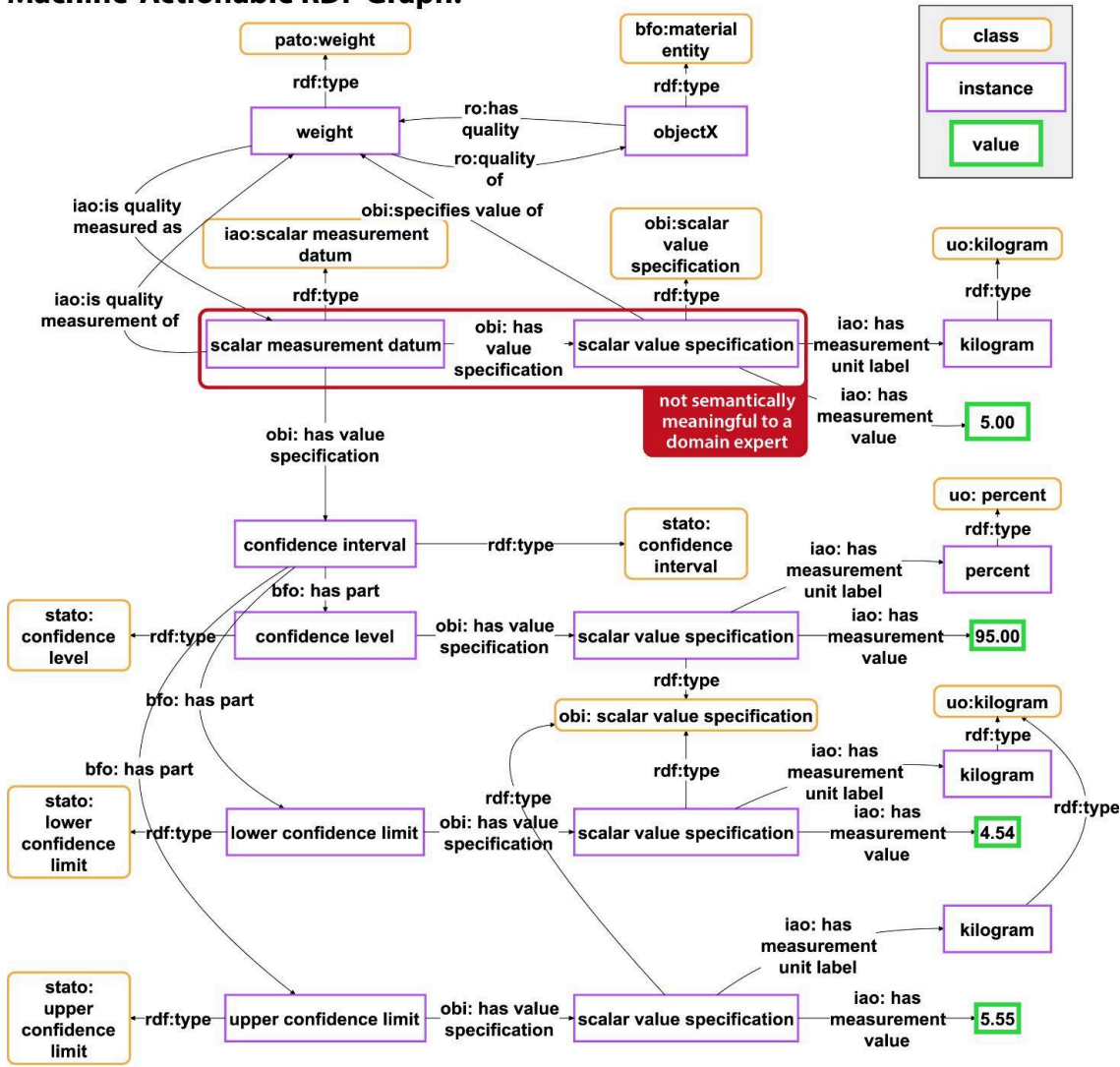

## Human-Actionable Mind-Map like Graph:

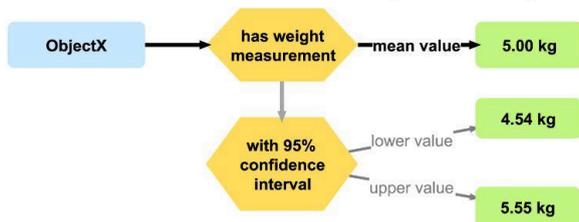

**Figure 1: Comparison of a human-readable statement with its machine-actionable representation as a semantic graph following the RDF syntax and with its human-actionable representation as a mind-map like graph. Top**: A human-readable statement about the observation that objectX weighs 5 kilograms, with a 95% confidence interval of 4.54 to 5.55 kilograms. **Middle**: A representation of the same statement as a graph, using RDF and following the general pattern for measurement data from the Ontology for Biomedical Investigations (OBI) (25) of the Open Biological and Biomedical Ontology (OBO) Foundry. Marked red is an example of a triple statement that is not semantically meaningful for a domain expert. **Bottom**: A representation of the same statement as a mind-map like graph, reducing the complexity of the RDF graph to the information that is actually relevant to a human reader.

# Challenge 2: Graph query languages are entry barriers for interacting with knowledge graphs

The typical KG is either a directed labeled graph that is based on RDF/OWL and stored in a tuple store, or it is a labeled property graph stored using, for instance, Neo4j. Directly interacting with the graph to add, search for, update, or delete data and metadata requires the use of a **graph query language**. For RDF/OWL, this is, for example, SPARQL, and for Neo4j it is Cypher.

Graph query languages allow detailed and very complex queries, but writing queries in SPARQL or Cypher is demanding. Someone without experience in writing such queries will have a hard time finding the data and metadata they are interested in. And many software developers are also not familiar with graph query languages and struggle with their complexity when attempting to learn them. Thus, having to write queries with a graph query language represents an entry barrier and therefore substantially limits the human-actionability of data and metadata stored in KGs (26).

Technical solutions to mitigate this issue are required, such as openly available and reusable query patterns that link to specific graph patterns. It also would be helpful if users could specify queries in a way that they look like mind-maps without having to use the formalization of a graph query language. The underlying semantic data schemata for such query graphs and their translation into SPARQL or Cypher queries, however, would have to be provided by the KG application and not by the users.

# Challenge 3: Knowledge graphs often do not support contextual exploration and visual information seeking strategies

In science and research, we are often dealing with a variety of 1-dimensional (e.g., lists of labels), 2-dimensional (e.g., geospatial data), 3-dimensional (e.g., 3-D models), temporal (e.g., timelines), and multidimensional (e.g., multi-attributes) data that can possess various inherent hierarchical structures (i.e., granularity trees) and cover different frames of reference (i.e., contexts). This diversity of data imposes specific requirements on the user interfaces (UIs) of information technology systems that have to facilitate complex tasks that users want to accomplish, including fact-finding, understanding cause-effect chains, or understanding controversial topics (27). Respective data exploration strategies should ideally promote not only **knowledge utility** (i.e., increase a user's domain knowledge) but also **exploration experience** (i.e., provide a user with a positive and pleasant exploration experience) and ultimately enhance **exploration effectiveness** (28). These requirements align with **visual information seeking mantras** such as:

1. ***Overview first, zoom and filter, then details-on-demand*** (29):
    1.1. Gain an overview of the contents of the entire information technology system,
    1.2. zoom in on data points and relationships of interest and filter out everything that is not of interest, and
    1.3. select a data point, a relationship, or groups of them and get details about them when needed.

2. ***Search first, show context, and expand on demand*** (30):
    2.1. Search for a specific data point,
    2.2. show relevant contexts given the user's current interests

2.3. so that users can expand contexts in the directions they are interested in.

3. ***Details first, show context, and overview last*** (31) ***or overview for navigation***:
   3.1. Start with a specific data point,
   3.2. show relevant contexts given the user's current interests
   3.3. so that users can explore these contexts in detail, using an overview for navigation purposes.

The *'Overview first'* strategy provides an intuitive guideline to the interaction requirements between the user's need for a broad awareness of the entire information space and all of its items (*seeing the entire collection* (29)) and the needs for seeing details, which is useful when dealing with datasets of moderate size (32). Conversely, the *'Search first'* strategy is optimized for large datasets, where providing a comprehensive overview for top-down analyses is often not feasible. *'Search first'* accounts for the needs of researchers who are not interested in general knowledge overviews or global patterns in the data, but seek answers to specific questions about one or several specific data points (32). The strategy is similar to online map search strategies, where search results provide the starting points for exploring local neighborhoods (33) and where *'context'* is understood as some sort of localized overview. The *'Details first'* strategy is designed for data such as spatial data, where experienced users know precisely where interesting data points lie and need support in exploring from there (31), with overviews serving as navigation aids.

In optimizing the explorability of information technology systems in general and KGs in particular, insights from all three mantras can be leveraged to enable zooming in and out of specific contexts and data points, facilitating information filtering, and providing overviews for effective navigation. These strategies contribute to enhancing the explorability capabilities of human users within the systems, promoting cognitive interoperability, and ultimately advancing data-driven research and analysis across domains.

Unfortunately, many KGs do not support any of these exploration strategies. Instead, KG users often explore the graph via a **semantic data browser**, which allows them to start exploring the graph from a single resource as the entry point, and moving from here along triple paths following RDF links (28,34). This approach is restricted to a single starting point, from which users explore the graph by navigating to individual candidate resources via triples that have the starting resource as their *Subject* and the candidate resources as their *Object*. For the graph shown in Figure 1, when using a semantic data browser, a user would have to click 15 times to gather all the information associated with the observation (Fig. 2). Obviously, this approach does not support any of the three exploration strategies mentioned above and makes access to information considerably more difficult, which reduces the overall (re)usability of data in a KG. Neither does providing KG visualizations that graphically display the contents of a KG, as these visualizations tend to also display irrelevant parts of the graph and do not scale well with a growing KG, already becoming overly complicated and information-dense for small sized graphs (35,36). However, approaches for KGs exist that support the three exploration strategies. According to (37), KG profiling and summarization, KG exploratory search, and KG exploratory analytics can be distinguished as main categories of general KG exploration approaches.

**Observation:**

*ObjectX weighs 5 kilograms, with a 95% confidence interval of 4.54 to 5.55*

**Number of clicks for gathering all information
in a semantic browser:**

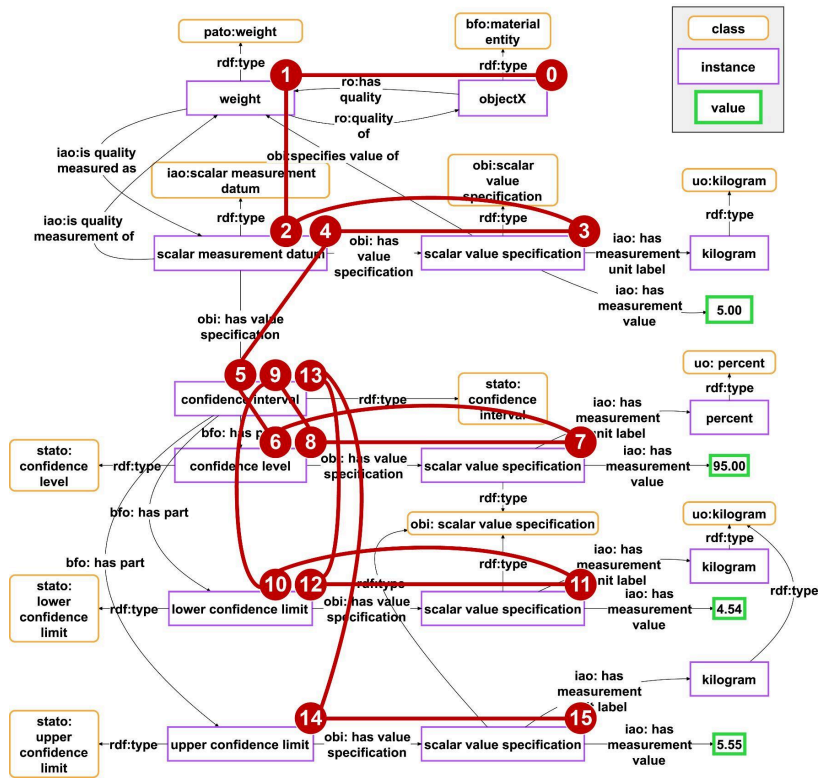

**Figure 2: Steps required for exploring the representation of an observation in a knowledge graph using a semantic data browser. Top**: A human-readable statement about the observation that objectX weighs 5 kilograms, with a 95% confidence interval of 4.54 to 5.55 kilograms. **Bottom**: A representation of the same statement as a graph (see also Fig. 1). The numbers in the red circles and the red lines indicate the number of clicking-step and the path that a user must follow to collect all the information about the observation in the KG, when using a semantic data browser.

**KG profiling and summarization** includes data profiling methods for computing basic statistics for a given dataset and make them available as descriptive metadata, providing information that supports users to get an overview of the dataset. In the context of KGs, this could be the number of classes and their instances or value distributions for specific properties (38). Some KGs also provide word clouds of resource labels, with font size indicating usage frequency, to support users in gaining a general overview (39). KG profiling and summarization also includes structural summarization (40,41) and pattern mining (42,43) to provide either a compact representation of the main features of the graph or create a somewhat simplified graph derived from the original graph. Compact representations and simplified graphs can be used to provide context to a given data point or an overview of the typical relationships of entities in the KG. KG profiling and summarization methods thus support users in gaining a **high-level overview of the graph** and therefore assist users in initial exploratory stages (37).

**KG exploratory search** refers to the open-ended exploration effort that combines browsing and searching for knowledge acquisition (44), often motivated by a vague information need (37). Exploratory search zooms in to the details of interest in a given dataset or graph, with the difference to traditional searches that the user first has to investigate how to search in order to find the data they are interested in. This behavior can be supported by approaches such as search-by-example (45,46) or domain-specific query interfaces that support users in specifying SPARQL queries without having to write SPARQL (47,48). Faceted exploration is a way to support KG exploratory search, allowing users to select facets as filters to decrease the size of the graph to what is relevant (49–51).

KG search interfaces that are based on facets can also be developed in such a way that they can be used for gradually formulating complex analytic queries without having to write SPARQL (52). KG exploratory search thus supports users in **identifying and exploring entities and relations of interest in the graph**, providing **contextual knowledge** that informs next exploration steps.

   **KG exploratory analytics** is an iterative process of data discovery and analytical querying of data which are not well known to the users (37). It requires providing multidimensional analysis functionalities over knowledge graphs that are typical of relational data warehouses, by describing multidimensional and statistical information within the KG (53,54). These approaches provide **specific views on the graph**, aggregating data into meaningful and relevant data collections tailored to **serving specific data needs**.

   Although these concepts facilitate contextual exploration of KGs, they are not sufficient. Moreover, they typically represent solutions that are tailored to a specific use case and cannot be directly transferred to other domains and use cases. We need innovative general exploratory methods and tools that enhance human explorability of data in a KG.

# FAIREr Guiding Principles

The effective utilization of KGs to enhance the machine-actionability and interoperability of data and metadata has garnered significant attention across various domains. However, the inherent challenge lies in striking a delicate balance between accommodating the needs of machines and catering to the needs and the cognitive capabilities of human users of KGs and thus domain experts, data scientists, and software developers. We argue that to address this challenge, a paradigm shift is imperative, necessitating extensions to both the EOSC IF and the FAIR Guiding Principles.

## The EOSC Interoperability Framework and the FAIR 2.0 Guiding Principles

According to the **EOSC IF**, achieving interoperability in practice requires the consideration of four dimensions, i.e., four distinct layers of interoperability (6). Information technology systems:

1. must work with other information technology systems in implementation or access without any restrictions or with controlled access for **technical interoperability**;
2. must provide contextual semantics related to common semantic resources for **semantic interoperability**;
3. must define contextual processes related to common process resources for **organizational interoperability**; and
4. must specify contextual licenses related to common license resources for **legal interoperability**.

   In the context of EOSC, a significant emphasis is placed on semantic interoperability, with the remaining layers of the EOSC IF serving its practical realization. Semantic interoperability plays a central role in enabling the effective exchange of data and metadata. This encompasses both human-machine and machine-machine interactions, aiming to ensure seamless and actionable communication. According to EOSC, semantic interoperability is achieved *"when the information* *transferred has, in its communicated form, all of the meaning required for the receiving system to interpret it correctly"* (6) (p. 11).

However, it warrants consideration whether the four layers of the EOSC IF and the four principles of FAIR and their associated criteria are sufficient in ensuring the effective communication of data and metadata between humans and machines. In (55), we argue that semantic interoperability does not only require interoperable **terms** but also interoperable **statements (i.e., propositions)**, and that the FAIR Guiding Principles must take this into account.

Regarding **terminological interoperability**, we must differentiate between terms that share both their meaning and referent/extension and are therefore **ontologically interoperable** and terms that only share their referent/extension but convey different meanings and are therefore only **referentially interoperable** (e.g., *Morning Star* and *Evening Star*, which both refer to the planet Venus but convey different meanings, or *Clark Kent* and *Superman*). **Entity mappings** are used to communicate ontological and referential interoperability relations between terms from different ontologies for their machine-actionability.

Regarding **propositional interoperability**, we must differentiate between statements that are based on the same logical framework and are therefore **logically interoperable** (e.g., they are both modelled in OWL and one can run a reasoner to check their logical consistency) and statements that are of the same statement type and are modelled using the same semantic data schema and are therefore **schema interoperable**. **Schema crosswalks** are used to establish schema interoperability between statements that use different semantic data schemata for the same type of statement. Based on this analysis of semantic interoperability, additional criteria for the four FAIR Guiding Principles have been formulated, resulting in the suggestion of **FAIR 2.0** (55).

## The EOSC Interoperability Framework and cognitive interoperability

Whereas FAIR 2.0 with its emphasis on semantic interoperability contributes to the machine-actionability of data and metadata, it does not necessarily also contribute to their human-actionability. Regarding human-actionability, it is important to know which actions hold significance for users when interacting with data and metadata. While users want to reliably and exhaustively find all data they are interested in within a KG and frequently also want to integrate data from different KGs, all of which requires the data's semantic interoperability, above all, users primarily seek to correctly comprehend the data—human-interpretability of data is a prerequisite for their human-actionability. Comprehending data goes beyond being able to correctly understand the meaning of individual data points, and includes also being able to contextualize them and explore further data points from any given data point in semantically meaningful ways. In essence, it involves understanding the interconnections between different data points. This intricate level of comprehension mandates data to possess cognitive interoperability, enabling users to extract meaningful insights effectively.

**Cognitive interoperability** is a critical characteristic of data structures and information technology systems that plays an essential role in facilitating efficient communication of data and metadata with human users. By providing intuitive tools and functions, systems that support cognitive interoperability enable users to gain an overview of data, locate data they are interested in, and explore related data points in semantically meaningful and intuitive ways. The concept of cognitive interoperability encompasses not only **how humans prefer to interact with technology (human-computer interaction)** but also **how they interact with information (human information** **interaction)**, considering their general cognitive conditions. In the context of information technology systems such as KGs, achieving cognitive interoperability necessitates tools that increase the user's awareness of the system's contents, that aid in understanding their meaning, support data and metadata communication, enhance content trustworthiness, facilitate integration into other workflows and software tools, and that clarify available actions and data operations. Additionally, cognitive interoperability also encompasses ease of implementation of data structures and their management for developers and operators of information technology systems. It thus addresses the specific data, tool, and service needs of the **three main personas** (56) identified for users of information management systems such as KGs, namely **information management system builders** (i.e., information architects, database admins), **data analysts** (i.e., researchers, data scientists, machine learning experts), and **data consumers** (i.e., stakeholders, end users, domain experts).

Knowledge representation should always accommodate cognitive limitations of human users, and data and metadata standards should support their cognitive interoperability. When developing UIs, for instance by including language-specific labels for multiple languages, including synonyms in data and metadata representations (55) and utilizing them in searches, considering the fact that humans can hold only 5-9 items in memory (57), and that search tasks can differ in complexity (e.g., fact-finding, understanding cause-effect chains, or understanding controversial topics) (27). As information technology systems grow in complexity and size, innovative exploratory methods and tools (37,58) become crucial to enhance **human explorability of data** and are important to improve the general interoperability between their data and the cognitive capabilities of human users. This can be accomplished by **reducing the complexity of the information displayed** in the system's UI to what currently interests a user, and by providing users semantically meaningful ways to access and explore the graph from any given data point. We thus argue that cognitive interoperability represents an essential aspect of the general interoperability of data and metadata, and we therefore suggest **adding cognitive interoperability as a fifth layer to the four layers identified by the EOSC IF**.

## Extending FAIR to FAIRER

*"Because human thoughts are combinatorial (simple parts combine) and recursive (parts can be embedded within parts), breathtaking expanses of knowledge can be explored with a finite inventory of mental tools."* ((59), p.360).

To address the challenges of achieving cognitive interoperability and overcoming the three challenges identified above (see *Problem statement*), an extension to the FAIR 2.0 Guiding Principles to include the **Principle of human Explorability** is introduced, leading to the formulation of the **FAIRER Guiding Principles** (i.e., **FAIR** + human **E**xplorability **r**aised; see Box 2), with the ultimate goal of striking a balance between accommodating the needs of machines and the needs of human users, and thus between the needs for machine-actionability and human-actionability of data and metadata. Whereas the FAIR Guiding Principles place a strong emphasis on the machine-actionability of data and metadata, the Principle of human Explorability does that for their human-actionability, drawing inspiration from fundamental principles of human cognition. Improving human explorability in FAIR information technology systems elevates their cognitive interoperability, turning FAIR information technology systems into **FAIRER information technology systems**.

Following the Principle of human Explorability, to support the abovementioned exploration strategies, data and metadata must be **organized into semantically meaningful subsets**, each uniquely represented by a GUPRI, enabling easy referencing and identification of the subset. Ideally, **each subset is organized as a FDO** and is classified as an instance of a semantically defined **FDO class**, documented in a corresponding ontology or controlled vocabulary (Box 2: **E1.**). Subsets can be **(recursively) combined** to **encapsulate complexity into manageable units**, forming subsets of coarser granularity (Box 2: **E2.**) which, ideally, are also organized as FDOs instantiating corresponding FDO classes. The organization of a KG into such subsets results in a significant increase in the human-explorability of data and metadata. Supporting human explorability of data also aims at enhancing the flexibility of knowledge management within information technology systems and overall enhancing the expressivity of such systems. Within KGs, this can be achieved by structuring the graph into separate, semantically meaningful subgraphs, which can be recursively combined to form larger subgraphs, with each subgraph being represented in the KG by its own resource and organized as a FDO.

In human communication, **propositions** are the smallest units of semantically meaningful information, conveyed through statements relating proper names and kind terms[2]. The RDF framework embodies statements as triples, with individual entities, concepts, and their connections represented as instance, class, and predicate resources, respectively. However, natural language statements often comprise more than one object and thus involve **n-ary relations**, which cannot be captured within a single triple, such as the statement *"Peter travels by train from Berlin to Paris on May 25th 2024"*. Therefore, RDF triples frequently map to natural language statements in a many-to-one instead of a one-to-one relation. Consequently, in a KG, simple statements featuring a single verb or predicate are modeled using one or more triples and constitute the smallest type of semantically meaningful subgraphs.

Understanding **human-actionable statements** and not single RDF triples as the **main units of communication** is essential when optimizing UIs for cognitive interoperability. Every datum in a dataset can be viewed as an individual statement, either explicitly or implicitly through its contextualization (e.g., as a list of measurement values in a column of a table that is interpreted as weight measurements with *kg* as their unit). By organizing triples that model a particular statement in their own subgraph, a KG can be organized into a set of subgraphs each of which maps to a particular statement (Box 2: **E1.1**). Such statement subgraphs can be organized into **nanopublications** (60–62) and would form statement FDOs.

A **statement FDO** represents a fundamental unit of information, comprising the smallest, most granular unit of information. It takes the form of a specific proposition, such as a volume measurement. Tabular data structures can be organized similarly by for instance structuring statements as single rows in the table, with each row having its own GUPRI that refers to the entire statement and that defines a corresponding statement FDO.

In addition to having its own GUPRI and the typical FDO metadata, each statement FDO should:

1. provide a specification of who created the FDO and distinguish it from who authored its content;

---

[2] Here, the focus is on the communication of semantic conceptual content and thus content organized and documented in the form of texts and symbols. Considering the communication of perception-based non-conceptual content, as when using media such as images, diagrams, audios, and videos, is out of the scope of this paper.

2.  reference the schema GUPRI for the data schema (i.e., [SHACL](#) shape, table structure, etc.) that was used for modelling the statement, to support schema interoperability (Box 2: F6.1) (55);

3.  specify the formal logical framework, if any, that was used for modelling the statement to indicate whether the FDO's content supports reasoning and which logical framework must be used, to support logical interoperability (Box 2: I5) (55);

4.  specify the statement category, distinguishing assertional statements (e.g., *Swan Anton is white*) from contingent (e.g., *Swans can be white*), prototypical (e.g., *Swans are typically white*), and universal statements (e.g., *Every swan is white*) (Box 2: F7); and

5.  provide a human-readable representation of the statement to increase its cognitive interoperability (Box 2: **E3**).

Ideally, statements can be directly negated without having to define corresponding class expressions, as it is required in OWL (Box 2: **E1.2**).

Fine-grained statement FDOs can then be combined in an information technology system to form coarser-grained **container FDOs** (Box 2: **E2**) that **organize statements into semantically meaningful collections of statements**, forming **respective coarser-grained FDO types**, which can be organized into [Research Object Crates](#) (RO-Crates) or similar technical implementations. Each such container FDO should instantiate a corresponding FDO class that specifies the type of FDO, and different container FDO types should be distinguished.

Organizing data and metadata into FDOs of various types leads to a nested structure of data and metadata, representing **different levels of representational granularity** (Box 2: **E2.1**). Additional types of subsets/subgraphs and corresponding FDO types can be identified based on **contextual differences** (different frames of reference; Box 2: **E2.3**) and inherent **hierarchies**, i.e., granularity trees present in the data, such as taxonomies and partonomies (Box 2: **E2.2**). Organizing a graph or a dataset into semantically meaningful subgraphs or subsets and structuring them into corresponding FDOs with accompanying metadata not only contributes to their cognitive interoperability and general reusability but also their **propositional interoperability** (55).

Besides these structural aspects of organizing data and metadata into different types of semantically meaningful FDOs, explorability also requires **adequate tools and easily understandable interfaces that provide operations that support users in pursuing different exploration strategies**. We need tools and interfaces that **decouple the human-readable display of data and metadata from their machine-actionable storage** (Box 2: **E3.**). The organization of data and metadata into different types of FDOs should thereby facilitate the development of UIs that support new ways to explore data and metadata, including **mind-map-like graphical** UIs in addition to **form-based textual** UIs (Box 2: **E3.1**). By reducing the complexity of data and metadata displayed in the UI to what is relevant for human comprehension, while disregarding information that is only relevant to machines (for an example, see highlighted triple in Fig. 1), mind-map-like displays enhance human-actionability of data and metadata. They can also provide graphical visual representations of complex interrelationships between various entities that are easier to comprehend than from their textual representations—just think of trying to understand a textual representation of a family tree as opposed to its graphical representation. Integrating textual and graphical displays within the UI further promotes user-friendly interactions with the data.

Moreover, information technology systems and their UIs should utilize the modular structure of the graph or dataset being structured into semantically meaningful FDOs and allow users to zoom in and out of data displays and provide contextual information for data points currently in focus for data exploration (Box 2: **E3.2**). The UIs should also support **making statements about statements** intuitively (Box 2: **E3.3**), as well as allowing users **to specify query graphs** using form-based or mind-map-like interfaces **without requiring knowledge of query languages** (Box 2: **E3.4**).

All these considerations result in the extension of the FAIR 2.0 Guiding Principles to incorporate the Principle of human Explorability, forming the FAIREr Guiding Principles, which offer significant potential to enhance cognitive interoperability of data and metadata and address existing challenges of information technology systems (for the specification of the FAIREr Guiding Principles, see Box 2).

---

**Box 2 | The FAIREr Guiding Principles that extent the FAIR 2.0 Guiding Principles (55) (in regular font) with the Principle of human Explorability (in bold font).**

**To be Findable:**
F1.   (meta)data are assigned a globally unique and persistent identifier
F2.   data are described with rich metadata (defined by R1 below)
F3.   metadata clearly and explicitly include the identifier of the data it describes
F4.   (meta)data are registered or indexed in a searchable resource
F5.   vocabularies are used that support terminological interoperability
    F5.1   terms with the same meaning and reference/extension are ideally mapped across all relevant vocabularies through ontological and referential entity mappings
    F5.2   terms ideally include multilingual labels and specify all relevant synonyms
F6.   (meta)data schemata are used that support propositional interoperability
    F6.1   the same (meta)data schema is used for the same type of statement or collection of statement types, and the schema is referenced with its identifier in the statement's metadata
    F6.2   (meta)data schemata for the same type of statement are ideally aligned and mapped across all relevant schemata (i.e., schema crosswalks)
F7.   (meta)data use a formalism to clearly distinguish between lexical, assertional, contingent, prototypical, and universal statements

**To be Accessible:**
A1.   (meta)data are retrievable by their identifier using a standardized communications protocol
    A1.1   the protocol is open, free, and universally implementable
    A1.2   the protocol allows for an authentication and authorization procedure, where necessary
    A1.3   the protocol is compliant with existing data protection regulations (e.g., General Data Protection Regulation, GDPR)
A2.   metadata are accessible, even when the data are no longer available

**To be Interoperable:**
I1.   (meta)data use a formal, accessible, shared, and broadly applicable language for knowledge representation
I2.   (meta)data use vocabularies that follow FAIR principles
I3.   (meta)data include qualified references to other (meta)data
I4.   vocabularies used by (meta)data provide human-readable ontological definitions and, where applicable, human-readable recognition criteria (i.e., operational definitions) for their terms
I5.   (meta)data specify the logical framework that has been used for their modeling (e.g., description logics using OWL or first order logic using Common Logic Interchange Framework)
I6.   see F7.
I7.(1-2)  see F5.(1-2)
I8.(1-2)  see F6.(1-2)

**To be Reusable:**
R1.   (meta)data are richly described with a plurality of accurate and relevant attributes
    R1.1   (meta)data are released with a clear and accessible data usage license
    R1.2   (meta)data are associated with detailed provenance
    R1.3   (meta)data meet domain-relevant community standards
    R1.4   metadata indicate the certainty level of the truthfulness of the semantic content of data

**To be Explorable:**

**E1.** (meta)data are structured into semantically meaningful subsets, each represented by its own globally unique and persistent identifier (ideally forming a FAIR Digital Object) that enables its referencing and its identification, and that instantiates a corresponding semantically defined (FAIR Digital Object) class

   **E1.1** ideally, each binary and n-ary proposition contained in the (meta)data is structured into its own statement subset (i.e., a statement FAIR Digital Object), forming a smallest unit of information that is semantically meaningful to a human reader

   **E1.2** ideally, negations can be expressed directly at the level of such statement subsets (i.e., a negated statement FAIR Digital Object)

**E2.** (meta)data subsets can be (recursively) combined to form compound subsets (i.e., container FAIR Digital Objects), each with its own globally unique and persistent identifier, instantiating a corresponding semantically defined (container FAIR Digital Object) class

   **E2.1** some (meta)data subsets (i.e., container FAIR Digital Objects) are organized into different levels of representational granularity

   **E2.2** ideally, each granularity tree contained in the (meta)data is structured into its own granularity tree subset (i.e., a granularity tree container FAIR Digital Object)

   **E2.3** ideally, each frame of reference contained in the (meta)data is structured into its own context subset (i.e., a context container FAIR Digital Object)

**E3.** human-readable display of (meta)data is decoupled from machine-actionable (meta)data storage to reduce the complexity of (meta)data displayed to human readers and to display only information that is relevant to human readers

   **E3.1** a user interface provides a form-based textual and ideally also a mind-map like graphical option for accessing and exploring (meta)data

   **E3.2** a user interface enables zooming in and out of (meta)data and provides contextual information by utilizing the different types of subsets (e.g., different types of statement and container FAIR Digital Objects)

   **E3.3** a user interface supports making statements about statements

   **E3.4** a user interface enables querying the (meta)data without requiring users to have knowledge of (graph) query languages

# Semantic units as FAIR Digital Objects: a strategy for Going FAIRer

Enhancing the cognitive interoperability of a FAIR KG to elevate it to the level of a FAIRer KG and to address the three challenges discussed above in the *Problem statement* requires a specific organization of the KG and its data (see FAIRer Guiding Principles, above). **Semantic units** (63,64) provide the required organization by structuring a KG into **identifiable sets of triples** and thus **subgraphs**. Unlike conventional methods of partitioning a knowledge graph (65–73), the semantic units approach focuses on structuring the graph into units of representation that are **semantically meaningful and easily comprehensible to human users of the KG** (Fig. 3) **[E1.]**.

Technically, a semantic unit is a subgraph within the KG. It is represented in the KG with its own resource, designated as GUPRI that **identifies** the associated subgraph **[E1.]**. Consequently, referring to a semantic unit resource via its GUPRI is equivalent to referring to the contents of its data graph, empowering KG users to make statements about the content encapsulated within the semantic unit's data graph **[E3.3.]**.

Organizing a KG into semantic units introduced a new layer of triples into the graph. We call the existing graph the **data graph layer**, and all newly added triples constitute the **semantic-units graph layer** of the KG. Each semantic unit resource, together with all triples that have a semantic unit resource in their *Subject* or *Object* position, belongs to the semantic-units graph layer. Analogously, we can distinguish a data graph and a semantic-units graph part for each semantic unit (Fig. 3A). Consequently, merging the data graphs of all semantic units of a KG results in the data graph layer of that KG.

Every semantic unit resource instantiates a corresponding **semantic unit class [E1.]**, which includes a human-readable description of the type of information covered by its instances and by their associated data graphs. Semantic unit resources introduce a new type of representational entity, **augmenting the expressivity of KGs** beyond instances, classes, and relations (i.e., RDF predicates). Semantic units can be accessed, searched, and reused as identifiable and reusable data items, forming units of representation that allow their implementation as FDOs in the form of nanopublications (63) or RO-Crates using RDF/OWL-based graphs (or any other technical implementation such as labelled property graphs or tabular data structures (64)).

Specifying **metadata** for a semantic unit is straightforward and should include creator, creation date, contributor, last updated on, author, and a copyright license specification (see Fig. 3A; the various metadata properties are indicated by *some metadata property* as a placeholder). Metadata triples for a semantic unit are located in the semantic-units graph layer.

### A) Statement Unit

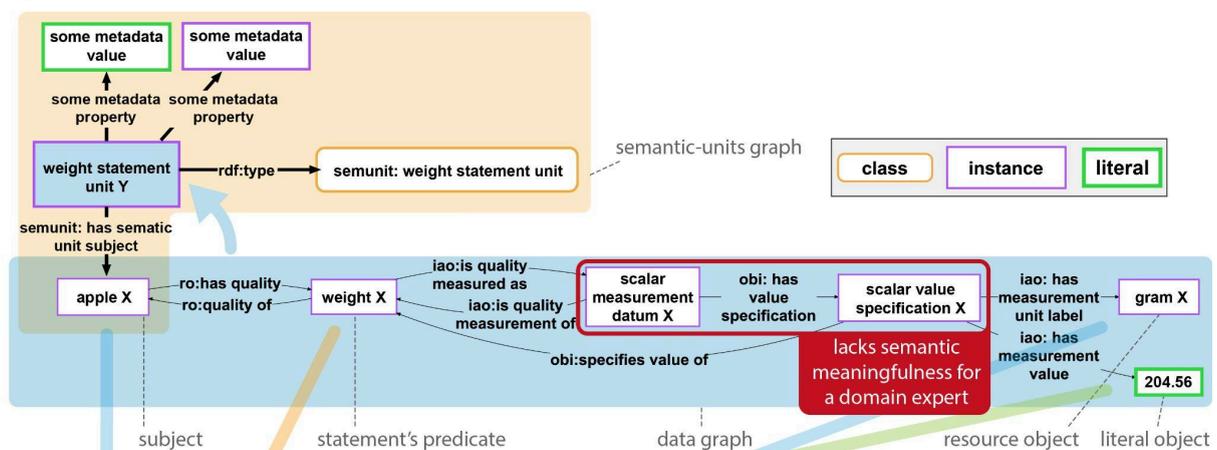

### B) Dynamic Label (Textual Display)

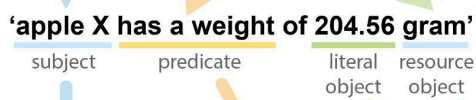

### C) Dynamic Mind-Map Pattern (Graphical Display)

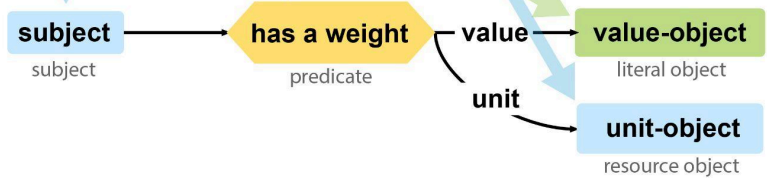

**Figure 3: Example of a semantic unit. A)** The statement *"Apple X has a weight of 204.56 grams"* modelled in RDF/OWL, organized as a statement unit, which is a specific type of semantic unit. The data graph, denoted within the blue box, articulates the statement with ‘apple X’ as its subject and ‘gram X’ alongside the numerical value 204.56 as its objects. The peach-colored box encompasses its semantic-units graph. It explicitly denotes the resource embodying the statement unit (bordered blue box) as an instance of the *SEMUNIT:weight statement unit* class, with ‘apple X’ identified as the subject of the statement. Notably, the GUPRI of the statement unit (*SEMUNIT:weight statement unit*) is also the GUPRI of the semantic unit's data graph (the subgraph in the not bordered blue box). The semantic-units graph also contains various metadata triples, here only indicated by *some metadata property* and *some metadata value* as their placeholders. Highlighted in red within the data graph is an example of a triple that is required for modelling purposes but lacks semantic meaningfulness for most domain experts. The dynamic label **B)** and the dynamic mind-map pattern **C)** associated with the statement unit class (*SEMUNIT:weight statement unit*). [Figure modified from (64)]

## Statement units

Different categories of semantic units can be distinguished (63,64). A statement unit represents the **smallest, independent proposition that is semantically meaningful for a human reader [E1.1]**. Depending on the aryness of the underlying statement, the data graph of a statement unit comprises one or more triples (Fig. 3A). Structuring a KG into statement units **mathematically partitions** the KG's data graph layer, ensuring that each of its triples belongs to precisely one statement unit.

By assigning to each statement unit class a specific graph schema that specifies how statement units of that type must be modeled, for instance in the form of a SHACL shape specification (74) that possesses its own GUPRI and includes constraints for its slots, semantic units also support **schema interoperability [F6.1;I8.1;I1.;I2.]**. By including a specification of the logical framework that has been applied (e.g., description logics or first order logic), **logical interoperability** can be specified for all statement units that are based on the same logical framework **[I5.]**.

Dynamic labels and dynamic mind-map patterns enhance the human-actionability of statement units. They are generated by parsing the label from the subject resource and the labels from the various object resources and literals of a given statement unit instance, and by creating from them a human-readable statement in the UI of a KG **[E3.1]** following a pre-defined template that is based on defined slots in a corresponding SHACL shape. Each statement unit class has its associated SHACL shape and templates for dynamic labels and mind-map patterns. **Dynamic labels** allow the generation of human-readable textual representations of statements, facilitating the display of complex n-ary propositions in a concise and understandable manner (Fig. 3B). For example, the template '*PERSON* travels by *TRANSPORTATION* from *DEPARTURE_LOCATION* to *DESTINATION_LOCATION* on the *DATETIME*' with the input '*Carla | train | Paris | Berlin | 29th of June*' would return the dynamic label '*Carla* travels by *train* from *Paris* to *Berlin* on the *29th of June 2022*'.

In the same way, **dynamic mind-map patterns** can be specified for the graphical representation of statements in the UI of a KG (Fig. 3C). They do not have to follow the RDF syntax of *Subject-Predicate-Object* and can thus directly represent n-ary relations without translating them into sets of binary triple statements. Moreover, both dynamic labels and dynamic mind-map patterns do not have to show all information in the data graph of a statement unit, but can restrict themselves to display only what is actually relevant to a human reader by ignoring all information that is additionally necessary for achieving the machine-actionability of the statement (Fig. 3: compare *A* with *B* and *C*). This results in a clear separation of machine-actionable storage from human-actionable display of data and metadata in a FAIRer KG **[E3.]**, which significantly enhances the content's cognitive interoperability. It allows exploring the contents of a FAIRer KG at the level of human-readable statements, where a statement that may be modeled in RDF with more than 20 triples could be represented in a single statement unit with a human-readable dynamic label or a less complex mind-map pattern, therewith reducing the complexity of the graph by filtering out information that is only relevant to machines and of no interest to a human reader (see *Challenge 1*). This also supports the development of UIs that understand statements as the main units of communication.

Statement units can be organized into **nanopublications** (63) (or any other technical implementation) and could provide a framework for **statement FDOs**. Several subcategories of statement units can be distinguished (see (64)).

By introducing some-instance, most-instances, and every-instance resources as new types of representational entities in addition to named individual (i.e., instance), class, relation (i.e., RDF predicate), and semantic unit resources, we can for instance distinguish lexical, assertional, contingent, prototypical, and universal statement units **[F7./I6.]**. **Lexical statements** are statements that use annotation properties and refer to linguistic entities. The other statement types can be distinguished by their subject resource, with **assertional statements** having named individuals as subjects, **contingent statements** some-instance resources, **prototypical statements** most-instances resources, and **universal statements** classes and every-instance resources as their subjects. Contrary to OWL, the semantic units framework allows representing universal statements and thus also class axioms without the use of blank nodes because they can be modeled and documented directly in a KG and thus become part of the domain of discourse of the KG (64) (for formal semantics see (75)).

In addition to this basic classification of statement units, they can also be classified based on the type of predicate they model (e.g., *has-first-name*, *has-part*, *derives-from*, *has-value*, *gives-to*, *travels-by-from-to-on-the*), resulting in a taxonomy of different statement unit classes. This taxonomy can contribute to a **general classification of corresponding statement FDOs**.

**Figure 4: Relation between an absence observation and two alternative ways of modelling it in a KG. A)** A human-readable statement about the observation that a given head has no antenna. **B)** The translation of the assertion from A) into an OWL expression mapped to RDF. Note how *absence phenotype* is defined as a set of relations of subclass and complement restrictions involving two blank nodes. **C)** The same statement can be modeled using two statement units. One of them is modeling the has-part relation and negates it (red box with purple borders, showing its dynamic label, with its data graph in the red box without borders). It is therefore an instance of *has-part statement unit*, *assertional statement unit*, and *negation unit*. The other semantic unit states that the resource 'some antenna' represents an antenna (UBERON:0000972), and its data graph is shown in the blue box. The combination of these two statement units models the observation from A). Utilizing the dynamic display patterns of the statement unit, the graph can be displayed in the UI of a knowledge graph application in a human-readable form, either as text through a dynamic label **E)** or as a mind-map through a dynamic mind-map pattern **F)**. *For reason of clarity of representation, the relation between 'head x' and head (UBERON:0000033) is not shown in B) and C).* [Figure based on figure from (64)]

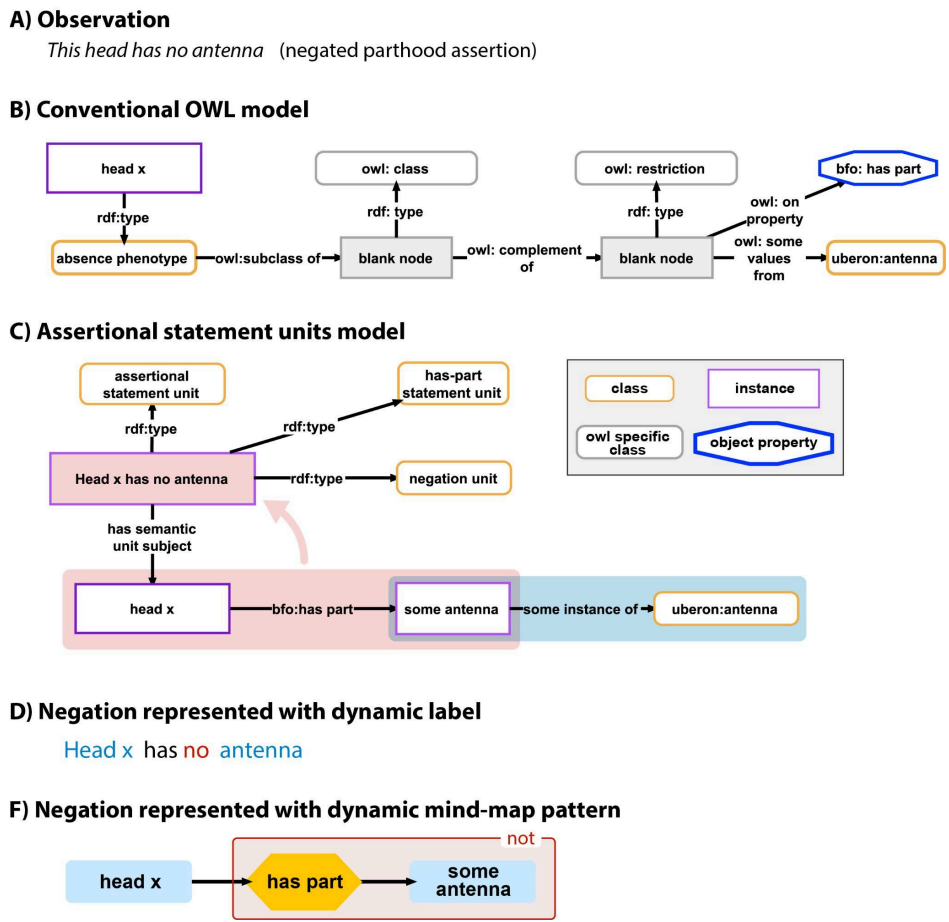

The semantic units framework also allows expressing **negations** (and cardinality restrictions) as instance graphs without using blank nodes and with a substantially simplified representation compared to OWL's axiomatic expressions (64,75), especially when using dynamic labels or mind-map patterns (see Fig. 4) **[E1.2]** (for formal semantics see (75)). Structuring a KG into semantic units thus substantially increases their overall expressivity.

## Compound units

Compound units facilitate the organization of statement units and other compound units into larger, **semantically meaningful** data graphs (Fig. 5) **[I3.;E1.;E2.]**. A compound unit can be viewed as a container of a collection of associated semantic units. Merging the data graphs of its associated semantic units constitutes the data graph of the compound unit. Analog to statement units, each compound unit is a semantic unit that is represented in the graph by its own resource and thus possesses its own GUPRI, and it instantiates a corresponding compound unit ontology class **[E1.]**. Compound units can be organized into RO-Crates (or any other technical implementation) and could provide a framework for **container FDOs and thus types of FDOs that are coarser-grained than statement FDOs**. Several subcategories of compound units can be distinguished, each serving a specific purpose in knowledge organization (63,64), resulting in a classification of compound units that can contribute to the **general classification of container FDOs**.

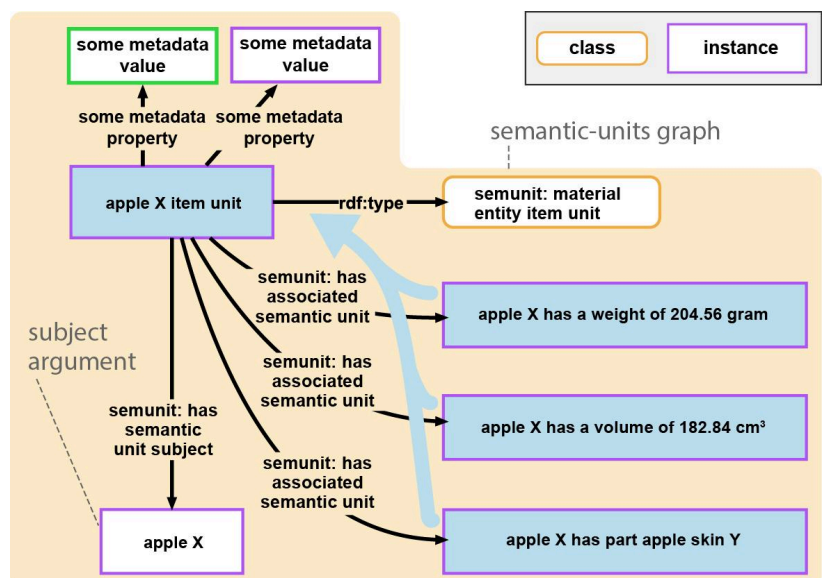

**Figure 5: Example of a compound unit** that comprises several statement units. Compound units possess only indirectly a data graph, through merging the data graphs of their associated statement units. The compound unit resource (here, *'apple X item unit'*), however, stands for these merged data graphs (indicated by the blue arrow). Compound units possess a semantic-units graph (shown in the peach-colored box), which documents the semantic units that are associated with it. [Figure taken from (64)]

An **item unit** (63) is a compound unit that associates all statement units of a FAIRer KG that share the same subject resource, and this subject resource also serves as the subject of the item unit **[E2.]**. For example, all statement units with the resource of a particular user as the subject would be associated with a profile item unit for that user, encompassing statements related to the user's username, email address, homepage, etc. Item units represent semantic units at the next coarser level of **representational granularity**, above the level of statement units **[E2.1]**. The information within an item unit is ideal for display on a dedicated page or knowledge panel in the KG's UI. Moreover, item units can be visually represented using **dynamic labels**, providing a concise yet informative display consisting of the label of their subject resource with some string added that is specific to the corresponding item unit class (e.g., *username*'s profile page) **[E3.1]**.

An **item group unit** (63) is a compound unit comprising at least two item units that are semantically linked through statement units sharing the same subject as one item unit and one of its objects as the subject of another item unit. Through such linking statement units, a chain of interconnected item units can be formed that can be represented in an item group unit **[E2.]**. Item group units represent semantic units at the next coarser level of **representational granularity**, above the level of item units **[E2.1]**. An item group unit is well-suited for presentation as a collection of interlinked UI pages, each page displaying the contents of a particular item unit. This approach is especially useful for describing for instance the relationships between the different parts of a material entity or the different steps of a recipe, resulting in a chain or tree of interconnected pages for easy user exploration.

A **granularity tree unit** (63) is a compound unit consisting of two or more statement units that collectively form a granularity tree (76–78) **[E2.2]**. Any type of statement unit that is based on a **partial order relation** such as parthood, class-subclass subsumption, or derives-from, can give rise to a corresponding granularity perspective (79–81), of which a particular granularity tree is an instance. Statement units associated with a granularity tree unit are interlinked following a tree-hierarchy, where one statement unit's object serves as the subject of another statement of the same statement type. Granularity trees can be found in diverse scenarios, such as partonomies of assembly parts in a car or anatomical parts in a multicellular organism, as well as taxonomies of term classes in ontologies.

A **granular item group unit** (63) expands on the granularity tree unit, including corresponding item units for each resource within the granularity tree unit **[E2.]**. Granular item group units thus comprise all item units whose subject resources are part of the same granularity tree unit.

A **context unit** (63) is a compound unit that incorporates all semantic units whose merged data graphs form a connected graph, where each resource and triple connects to every other resource through a series of triples within the context unit's data graph, excluding **is-about statement units**, i.e., statements with *isAbout* (IAO:0000136) as their property that relate an information artifact to an entity that the artifact contains information about. The presence of is-about statement units in a graph thus indicate changes in **reference frames** within the graph **[E2.3]**.

A **standard information unit** (64) is a compound unit with a collection of statement units that form a well established or standardized piece of information such as a standard material data sheet in materials science, a product data sheet in industry, or the standardized information publicly displayed on a user profile page of a social media application **[E2.]**. A standard information unit comprises a defined collection of statement units that qualifies as a standard unit of information in a specific context or domain. Any standardized description of an object that requires a specific set of observational statements can be represented as a specific type of standard information unit. Standard information units can also be used to define domain- and topic-specific views on the knowledge graph.

A **logical argument unit** (64) is a compound unit that relates three statement units to each other to form a logical argument, with two statement units functioning as the premises and the third statement unit as the conclusion **[E2.]**. Depending on which statement unit types form the premises and the conclusion, deduction, induction, and abduction units can be distinguished.

Lastly, a **dataset unit** denotes a compound unit that defines an ordered collection of particular semantic units **[E2.]**. The GUPRI of a dataset unit aids in identifying and referencing arbitrary collections of semantic units not covered by other semantic unit categories, such as datasets imported to a KG or all semantic units contributed by a specific institute. It also allows users of a KG to define their own dataset units, which they can reference using its GUPRI.

## Question units

With the introduction of some-instance, most-instances, and every-instance resources, questions can be represented in a KG in the form of question units (64). A question unit utilizes one or more existing semantic unit classes as their **source**, creating instances of them as questions that can be documented as a connected subgraph in the KG, akin to other semantic units (Fig. 6).

Based on the SHACL shape specification of the statement unit classes involved, a query-builder can derive corresponding graph queries that can be executed by the KG. If the question unit leaves none of its subject and object positions underspecified, the question assumes a Boolean *true/false* answer. However, by using some-instance and every-instance resources as variables for subject and object resources and datatype specifications with underspecified values as variables for literal objects, respective queries will return a list of matching subgraphs (Fig. 6). For instance, when using a some-instance resource of the type underline{material entity} (BFO:0000040) as the subject resource in a question unit for has-part statement units, the query would return all has-part statement units that have some material entity as their subject. In the UI of a KG, the same forms for adding statements units can be used for adding question units.

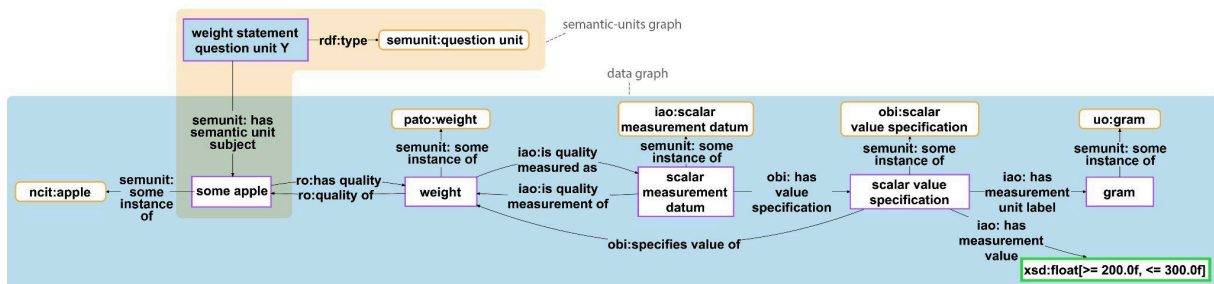

**Figure 6: Question units**. Example of a question unit. *For reasons of clarity, metadata for each semantic unit is not represented.*

Question units require a query-builder that automatically translates question units into graph queries in reference to the SHACL shape specification of the corresponding source statement unit class. The implementation of question units that can be represented in the KG and a corresponding query-builder effectively converts searches into objects within the KG, a strategy that bears resemblance to previous works (82,83) in the field.

Overall, the introduction of question units together with corresponding UIs and a query-builder would not only provide an intuitively usable framework for specifying SPARQL or Cypher queries without having to know graph query languages (see *Challenge 2*) **[E3.4]**, but could also support ontology development by documenting specific competency questions in the graph (84) in the form of a set of corresponding question units.

## Making statements about statements

The modular and nested organization of semantic units and the fact that each semantic unit in a KG can be represented as a FDO with its own GUPRI opens up new possibilities for making **statements** **about statements [E3.3]**, introducing an additional layer of communication and metadata enrichment to KGs. The representation of semantic units as FDOs with their own GUPRIs facilitates straightforward statement-making about them, with statement units as smallest units of information providing a fine-grained basis and various types of compound units as collections of statements more coarse-grained semantically meaningful units of information. Statements about semantic units can themselves be represented in a KG as semantic units and can encompass various aspects, such as associating provenance information with semantic units, indicating compliance with data protection regulations **[A1.3]** based on, e.g., GDPR, or specifying the logical framework used for data modeling **[I5.]**. Additionally, semantic units being organized as FDOs enable making statements across different databases and KGs, facilitating data sharing and cross-referencing of information (75).

Structuring a KG into semantic units and thus semantically meaningful units of information, each possessing its own GUPRI, also fosters a communication layer within FAIRer KGs, empowering users to express their own viewpoints and opinions in a structured and machine-actionable manner by adding statements about them to the graph.

To realize the full potential of makings statements about statements and leveraging its benefits effectively, the development of intuitive UIs becomes crucial. By structuring KGs into semantic units, UIs can be designed to facilitate making statements about statements. The adherence to a clear and straightforward schema and structure in implementing semantic units further streamlines the development of accompanying UIs, tools, and services.

# How semantic units make FAIR knowledge graphs FAIREr

## Semantic units structure a knowledge graph into different levels of representational granularity, granularity perspectives, and frames of reference

The semantic units approach provides a powerful organizational framework for enhancing a KG's overall structure and improving its semantic and cognitive interoperability. By organizing the data graph of a FAIR KG into identifiable and semantically meaningful units, a semantic-units graph layer is created that functions as the **discursive layer** of the KG that comprises **five levels of representational granularity** (63). At the finest grained level are individual triples, followed by statement units, item units, item group units, and the KG as a whole (Fig. 7) **[E2.1]**.

The use of semantic units facilitates the visualization of statements in a **mind-map-like** manner or as **natural language statements**, greatly aiding human comprehension and exploration of the KG's contents **[E3.; E3.1]**. All statements belonging to an item unit can be displayed on the same UI page, and the item units belonging to the same item group unit can be represented as a collection of semantically interrelated UI pages.

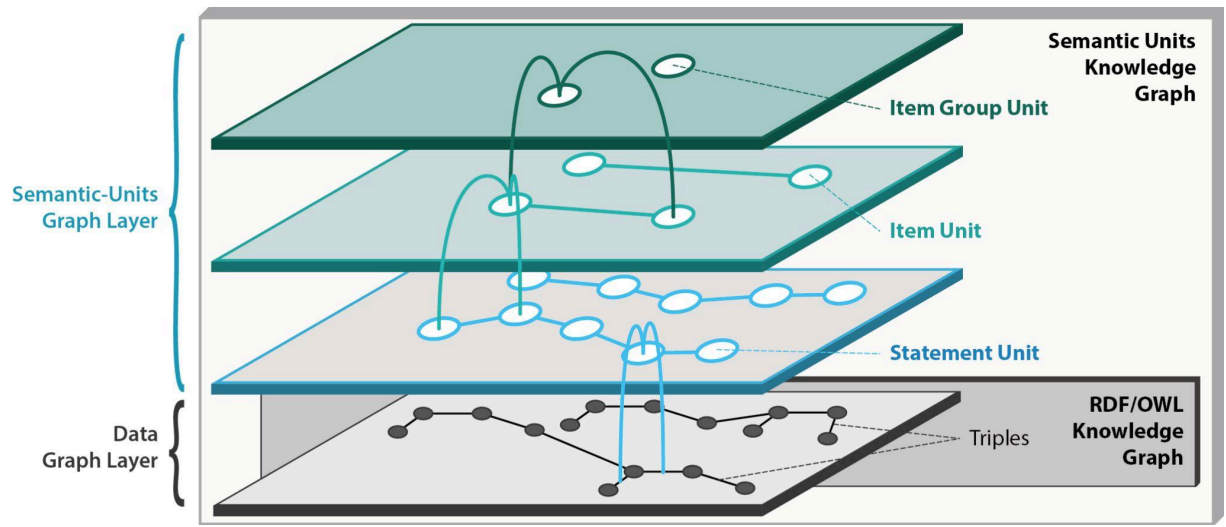

**Figure 7: Five levels of representational granularity.** The introduction of semantic units to a knowledge graph adds a semantic-units graph layer to its data graph layer, which adds a level of statement units, a level of item units, and a level of item group units to the level of triples and the level of the graph as a whole, resulting in five levels of representational granularity.

The KG's organization is further enriched by the introduction of granular item group units, achieved through the utilization of granularity tree units present in the graph. **Granularity trees organize the data graph layer and thus the ontological layer of a KG into different granularity perspectives, each with its own levels of granularity**, providing users with different viewpoints to explore and analyze the KG's contents effectively. Granularity tree units organize the KG orthogonally to and independent of representational granularity. In other words, while representational granularity organizes the way *how* ontological knowledge is communicated (i.e., the discursive layer), granularity perspectives organize the ontological knowledge itself (i.e., the ontological layer). Finally, context units distinguish different **frames of reference** across the ontological and discursive layers of the KG.

The organization of a KG into semantic units and thus semantically meaningful, partially overlapping and partially enclosed subgraphs, lays the foundation for the development of advanced graphical UIs with innovative information-exploration tools, that enable users to access data and metadata in new ways and to rapidly explore them in a user-controlled fashion, with visual representations of data and metadata, thus supporting **visual information seeking strategies** mentioned above (*'Overview first'*, *'Search first'*, and *'Details first'*). These UIs should act as **magic lenses**, enabling users to find, sort, filter, and present relevant information in user-controlled, intuitive ways.

By leveraging the organization of data and metadata provided by FAIRer KGs, users can swiftly gain an **overview** of the KG's contents by exploring the graph at the level of item group units, with the possibility to **zoom in** on finer levels of representational granularity, granularity trees, or contexts **[E3.2]**, thereby **filtering out** data and metadata they are currently not interested in and rapidly **getting details-on-demand**. Granularity tree units with their intrinsic hierarchical organization each structure the data graph of a FAIRer KG into different levels of granularity that, together with the representational granularity of statement, item, and item group units, provide **overviews for navigation purposes** and possibilities for **identifying contexts of interest to expand upon in directions a user is interested in** (see UI tool examples further below for).

Let us apply this to the representation of the scientific findings of a scholarly article in a FAIREr scholarly KG. As an example, we use a fictional article that describes the anatomical organization of a particular specimen in the form of a hierarchy of has-part relations between anatomical entities, forming a partonomy. Additionally, the description includes shape specifications, length and width measurements, and specifications of positional relationships for some of these anatomical entities (Fig. 8).

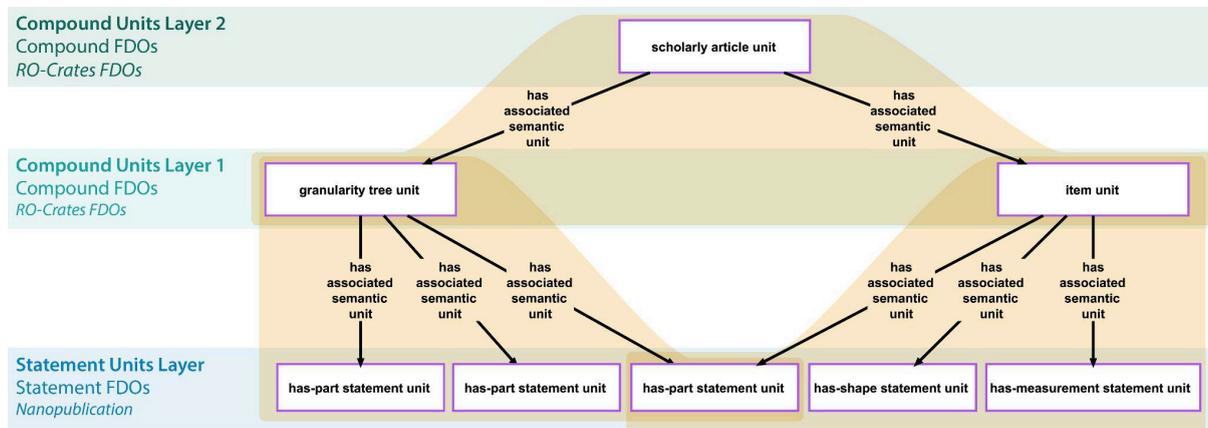

**Figure 8: Example of different layers of granularity in a scholarly FAIRer knowledge graph.** The scholarly FAIRer knowledge graph models scientific findings of different scholarly articles. Each article is modelled in the graph using different types of semantic units as FDOs. For an article describing the anatomical organization of a particular specimen, the description can be organzied into a layer of statement units at the bottom, a layer of compound units consisting of item units and a granularity tree unit in the middle, and a layer of different scholarly article units at the top. More layers can be added depending on the type of information modelled from the scholarly articles.

The phenotype description can be modelled in the KG as a set of statement units of the types *has-part*, *has-shape*, *has-length*, *has-width*, and *has-positional-relationship-to*, together forming a connected subgraph within the KG. Each of the statement units is organized as a nanopublication and represents a statement FDO. This allows researchers to individually refer to the smallest semantically meaningful units of information of the article, and to make statements about them within the KG.

All statement units that have the same anatomical entity resource in their subject position form an item unit. Each such item unit documents all information from the article that describes that particular anatomical part of the specimen. Organized as for instance an RO-Crate, each such item unit is a FDO that researches can refer to externally and make statements about within the KG.

The partonomy that describes the anatomical organization of the specimen is modelled using interconnected *has-part* statement units. The collection of these statement units is organized in the FAIREr KG as an RO-Crate and thus a FDO that is a granularity tree unit. Again, researchers can refer to the partonomy externally and make statements about it within the KG.

Finally, the entire subgraph representing the article in the FAIREr KG is organized as an RO-Crate that is a standard information unit of the type scholarly article unit. As a FDO, researchers can again refer to it and make statements about it.

When modelling the scientific findings of several scholarly articles in the FAIREr scholarly KG, the graph will be organized hierarchically into several scholarly article units that each comprises several item units and a granularity tree unit based on a particular collection of *has-part*, *has-shape*, *has-length*, *has-width*, and *has-positional-relationship-to* statement units. This would organize the graph into five distinct layers, from triples to statement units, compound units that are item and granularity tree units, compound units that are scholarly article units, and the graph as a whole (Fig. 8).

## Semantic units enhance data access and data exploration

In their search for specific data and metadata, users of KGs are often confronted with the task to find the needle in the haystack, especially when dealing with large and complex KGs. Finding specific information within the vast sea of data can be challenging, and effective data structures and supportive UIs are essential to support the users' needs. To this end, two main data exploration tasks must be distinguished based on different user motivations for accessing data and metadata in a FAIRer KG (85):

1. **Known-item search**: Users have a well-defined information need and a clear idea of the expected results (*conventional search* (86)), and aim to find a specific set of resources or relations in the KG that satisfy this need.

2. **Browse**: Users seek to develop a general understanding of the KG's contents or discover unexpected patterns within it. Such **exploratory search** requires an exploration effort and is usually open-ended, has unclear information needs, and is often used for learning and exploration tasks (28) (see also *challenge 3*). While exploring a KG using a semantic data browser following RDF links facilitates path-based exploration, it frequently poses challenges and confusion for users because of the high complexity of machine-actionable data representations, with many added resources that are not relevant for a human reader to understand the meaning (see *Challenge 1*; Fig. 1 Middle). The larger the KG becomes, the likelier it is that even domain experts find its contents to be increasingly hard to grasp, requiring exploratory tools to support them in comprehending the contents. In the past, however, the attention was more on KG construction than on the investigation and visualization of the graph itself, and consequently not many tools supporting KG exploration have been developed (28). While early approaches of facilitating exploratory searches focused on textual or visual interfaces (87,88), new approaches include the development of algorithms for generating exploration paths for knowledge expansion based on path parameters. Path parameters include the level of generality of a resource in focus and the direction of exploration that a user pursues within the class hierarchy, the density of entities forming an exploration path (i.e., number of direct connections of a given entity), and the amount of highly inclusive and familiar concepts (i.e., knowledge anchors) along the path. The algorithms attempt to identify path parameters that affect the utility of the exploration, with the overall goal to support meaningful learning (28). It has been shown that successful implementations of exploratory search strategies expand a user's domain knowledge when exploring unfamiliar domains (i.e. knowledge utility) (34,89).

In practice, users often combine these two tasks, because they desire to continue their exploration of the KG from their initial access point in semantically meaningful ways, using general overviews and relevant contextual information to facilitate efficient navigation.

FAIRer KGs introduce novel data access and semantically meaningful exploration points, improving the way users interact with and extract information from KGs. We outline four key data access and exploration points facilitated by FAIRer KGs and thus KGs that are organized into different types of FDOs, each of which corresponds with a specific type of semantic unit (see *Challenge 3*):

## 1. Exploring by ontology class

KGs usually organize class-terms into a taxonomy, enabling users to browse this taxonomy and trigger predefined Cypher or SPARQL queries by selecting specific terms. This feature allows users to identify all instances of a selected term within the graph—like a sophisticated **keyword search**. In FAIRer KGs, the keyword search would return a list of different types of FDOs, with each item in the list indicating the number of FDO instances that reference the respective class-term. Users can then decide which levels of representational and other types of granularity they are interested, enabling **zooming in, filtering out, and providing detail-on-demand** in a more structured and semantically meaningful way than conventional KGs do. **Each semantic unit provides a defined view on a specific part of the KG**, providing users with more options for data exploration.

Additionally, since every semantic unit instantiates a corresponding semantic unit class, one could also search and explore the KG using a **dynamic facetted search**, allowing users to filter search results using semantically relevant facets like FDO classes, in the case of statement units filter by resource types and value ranges and patterns for each slot of the corresponding SHACL shape, and for compound FDOs by a list of associated statement FDO classes.

By enriching the taxonomy of class-terms with statistics about the number of their instances and the number of their mentionings in statement, item, item group, granularity tree, and context FDOs, coupled with the possibility to choose between different time intervals of their creation or last-updated date (e.g., within last week, last month, last year, last decade), FDOs that correspond with semantic units support the **identification of hot-spots** within FAIRer KGs at various levels of granularity, providing users with additional **overview** options.

## 2. Exploring by semantic unit class

Each FDO in a FAIRer KG is represented by its own GUPRI and instantiates a corresponding FDO class. The set of all FDO classes forms a taxonomy, a nested hierarchy, with specialized FDO types being subclasses of more general parent FDO types. This **taxonomy of different FDO types and their corresponding semantic units** provides a basic classification for the contents of FAIRer KGs, with each individual FDO instantiating at least one corresponding FDO and semantic unit class. Users can access data by browsing the list of these classes. The ability to differentiate FDO classes based on subject and object resources, value ranges, and in the case of compound FDOs their types of associated FDOs, allows users to identify the specific types of FDOs they seek. Predefined Cypher and SPARQL queries can then retrieve the instances of respective semantic unit classes, and users can access the information at the level of representation, the granularity perspective, and the frame of reference they are interested in and filter it according to their preferences using dynamic facets. The different types of FDOs can also be used for **filtering** the contents of a KG, simplifying the KG's complexity by hiding irrelevant or highlighting relevant information.

Combining different types of FDOs and their corresponding semantic units in a search enables users to perform more comprehensive searches and access data at multiple levels. Users can, for instance, search for all instances of a specific type of statement FDO that are part of item FDOs that have a specific type of subject resource—e.g., all weight measurement statement FDO in phenotype description item group FDOs. Users can start with searching for instances of a specific ontology class, and after having selected a particular instance, the FAIRer KG application can indicate which statements, items, item groups, granularity trees, datasets, questions, or which frames of reference this resource is part of. This assists users in **identifying relevant contexts**, indicating semantically meaningful opportunities to further explore the KG from a given data point. A simple click can start a predefined Cypher or SPARQL query, and the data graph of a particular granularity tree or frame of reference that contains the previously selected resource can be shown in the UI.

FDOs that correspond with semantic units can also be utilized for offering a summary of the contents of FAIREr KGs at different degrees of generalization and abstraction by indicating the number of instances of each FDO class. This summary aids in **profiling a FAIREr KG** and gaining a general **overview** of its content. The human-readable descriptions associated with each statement FDO class in a FAIREr KG offer further valuable guidance for users exploring the KG. These descriptions provide insights into the type of information covered by a given type of FDO and can be displayed as tooltip texts in a UI, enhancing the user experience whenever the respective class or one of its instances is presented.

Implementing the concept of semantic units as FDOs in KGs will thus contribute to a solution to all three challenges, that of overly complex machine-actionable graphs (*Challenge 1*), the challenge that many users and even software developers do not know query languages and do not want to learn them (*Challenge 2*), and the challenge of supporting contextual exploration and visual information seeking strategies in KGs (*Challenge 3*).

### 3. External access through GUPRI

The GUPRI associated with each resource in a FAIREr KG allows external applications to refer to these resources unambiguously. Providing corresponding APIs enables external access to the FAIREr KG's resources, including instances, classes, and semantic units. As FDOs, semantic units can also offer provenance data and contextual metadata, facilitating data integration with external sources.

### 4. Tabular presentation

Statement units of the same type hold information of the same type, modeled according to the same shape. Therefore, their data are comparable and interoperable, making it feasible to present and visualize them in a single table, with each slot of the corresponding SHACL shape defining a column in the table and each statement unit instance a row.

Item units with the same type of subject resource tend to have similar associated statement units. By aligning statement units across item units, a single table can visualize the contents of multiple item units or item group units, with statement units being mapped to the rows and the item units to the columns. The results of queries can be visualized using a similar tabular approach.

## Semantic units enable novel user interfaces for graph exploration in FAIREr Knowledge Graphs

FAIREr KGs present a promising framework for developing innovative UIs that support efficient **navigation and exploration**, aligning with the **visual information seeking strategies** of *'Overview first'*, *'Search first'*, and *'Details first'*.

Apart from offering diverse access points for search and exploration (see above), UIs can incorporate a navigation tree for seamless browsing of contents related to a specific item group FDO or granular item group FDO **[E3.1;E3.2]**. This navigation tree builds on the hierarchy of interconnected item FDOs within the same (granular) item group FDO, which can become rather deep, with chains of multiple item FDOs connected via statement FDOs. By representing this navigation tree in the UI in a folder-like system, utilizing the human-readable dynamic labels of the corresponding semantic units, users can intuitively jump from one item FDO to another by a single click (Fig. 9). Since item FDOs can be linked to each other via different types of statement FDOs, one can even allow users to filter for specific types of linking statement FDOs. Consequently, a user can for instance decide that they only want to see item FDOs that are linked to each other via adjacency statement FDOs and the navigation tree would show the respective item FDO tree(s).

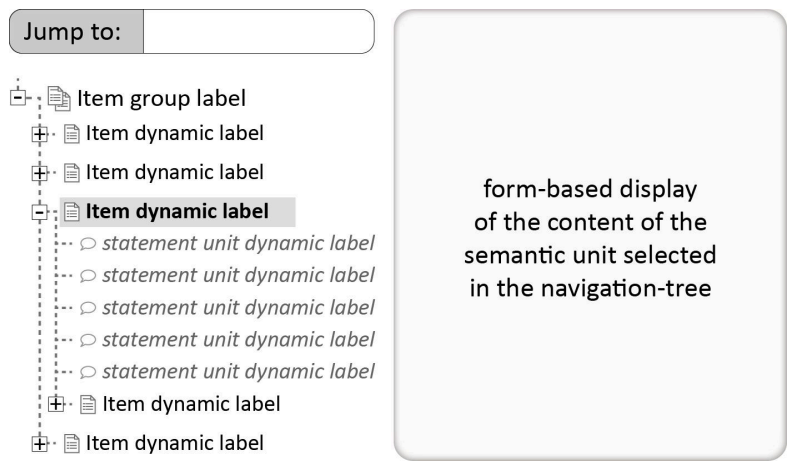

**Figure 9: Navigation tree for navigating and exploring data of a specific (granular) item group unit.** The tree to the left represents the hierarchy of a specific (granular) item group unit with all its associated item units, with 'child' item units describing in more detail a resource that their 'parent' item unit referred to as an object of one of its statement units. Users can use this tree for navigating the contents of the item group unit, zooming in and filtering out information, and getting more details on demand. The information from the selected item unit will then be displayed in a form, e.g., in a widget to the right. This way, users can have the items they are interested in expanded within the navigation tree and can easily jump between them by a simple click. Each semantic unit can be shown with its dynamic label. In the figure, the tree also includes statement units belonging to an item unit as leaves in the tree.

To demonstrate this navigation tree functionality, we have developed a corresponding UI in a Python-based prototype of a FAIREr scholarly KG application (90) (Fig. 10). The prototype exemplifies the core concept, though it lacks the ability to select different types of linking statement units. Additionally, the prototype includes versioning for semantic units, automated tracking of editing histories and provenance information. It is available from https://github.com/LarsVogt/Knowledge-Graph-Building-Blocks.

Often, **graphical representations in the form of mind-maps** are better suited for understanding relationships between entities of interest than textual representations **[E3.1;E3.2]**. However, conventional KGs often suffer from information overload, typically containing a lot of information that is only relevant to a machine, rendering the graph unnecessarily complex and less comprehensible for human readers (see, e.g., Fig. 1 Middle; *Challenge 1*). With the introduction of dynamic mind-map patterns and the distinction of representational granularity levels, FAIREr KGs allow for **filtering** the data graph into semantically meaningful subgraphs. The dynamic mind-map patterns **present only information that is relevant to users**, reducing the complexity of the underlying data graph to easily digestible chunks for a human reader. Users can **zoom in and out** across levels, traverse granularity trees, and explore different frames of reference in a manner that aligns with their interests. Users can, for instance, start to explore the graph with a search result that shows an item group FDO that contains information they searched for and all relations of this FDO to other item group FDOs (e.g., an observation-statement from publication *A* supports a hypothesis-statement in publication *B*). Moreover, all associated statement FDOs of a given item FDO could be shown and thus information from the next finer representational granularity level (Fig. 11).

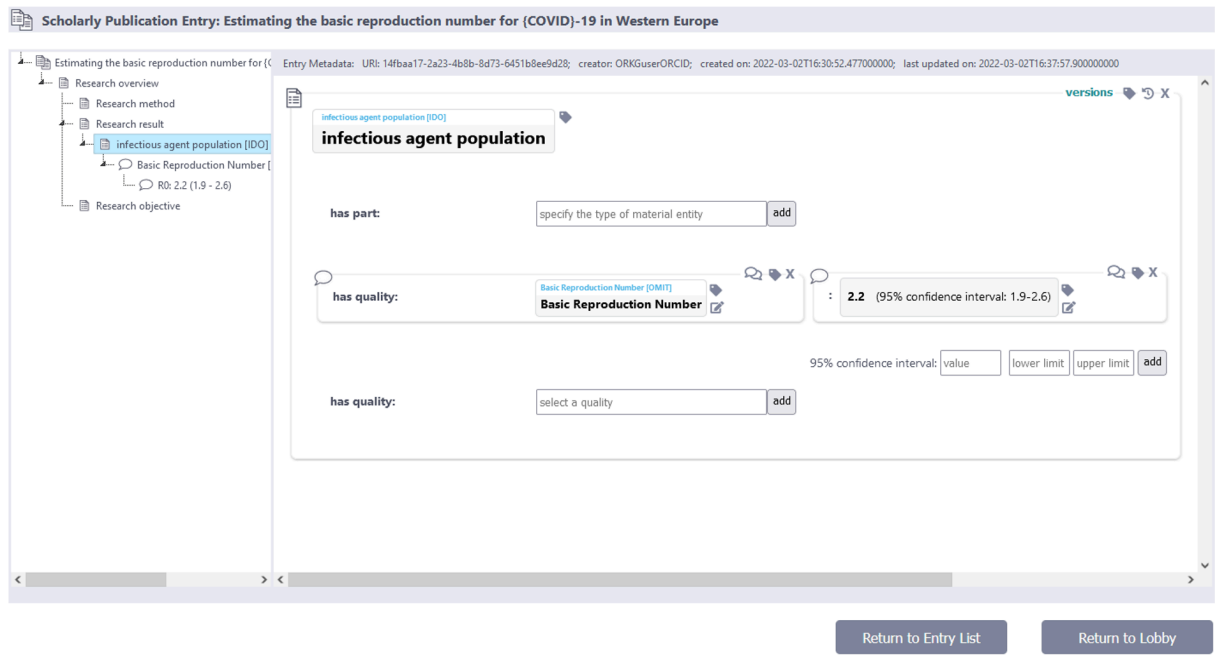

**Figure 10: UI showing a scholarly publication item group unit**. The left widget shows a navigation tree with the item group unit at the root (i.e., top) and all associated interlinked item units organized below in the tree. The content of the selected item unit is displayed in the widget to the right, using Jinja templates. The item unit describing *infectious agent population* (IDO:0000513) is selected, and it has two associated statement units that specify the population's basic reproduction number. Input fields allow adding more data.

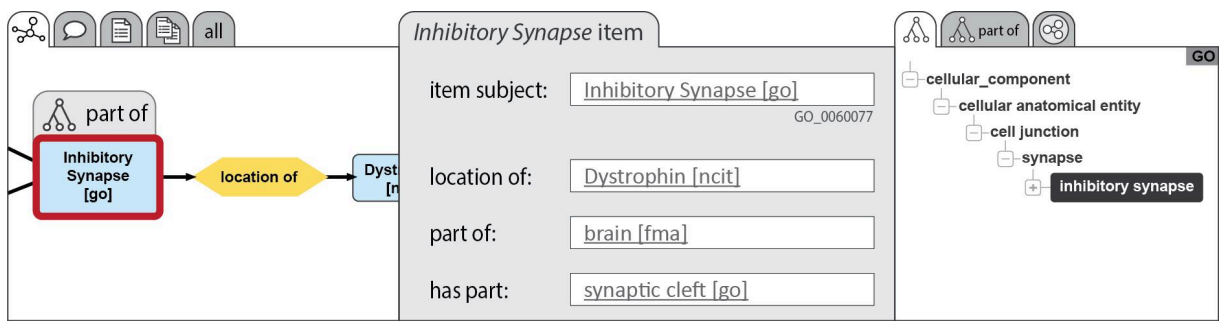

**Figure 11: Concept study of a UI with interacting mind-map like graphical and form-based textual widgets for knowledge graph exploration.** The UI is divided into three parts: to the left, the mind-map like graphical exploration widget for zooming in and out across all five levels of representational granularity, indicated by the tabs at the top left (triples, statements, items, item groups, the whole graph). Currently, the triples level is shown, and the 'inhibitory synapse' (GO:0060077) node is selected. This graphical UI allows users to explore the graph, select a node or a relation and, by clicking on one of the tabs above, switch to their representations at different levels of representational granularity. To the right, the graphical representation of the taxonomy for the currently selected resource is displayed, with tabs allowing to switch to other types of granularity trees (e.g., a parthood granularity tree) or frames of reference. The form-based textual widget in the middle shows the input form for the *'inhibitory synapse item unit'* and allows users to change, delete, or add new statement FDOs, with the other two widgets being updated automatically. When a user selects a relation in the left widget, the middle widget will show the corresponding statement FDO with its associated provenance information and the right widget the taxonomy of statement FDOs. Clicking into an input field in the form widget in the middle will select the corresponding subgraph in the graph widget to the left. All mind-map like displays in the graph widget to the left do not show the triples as they are stored in the machine-actionable data graph, but utilize the dynamic mind-map patterns that only consider the information that is relevant to a human user, guaranteeing the human-actionability of the data and metadata.

The identification of hierarchical structures like granularity trees and frames of reference via context FDOs empowers UIs of FAIREr KGs to facilitate **smart and semantically meaningful exploration**. By **displaying relevant contexts**, users can effortlessly **expand the graph in the direction** **of their interests**. The graph-exploration tools of UIs of FAIRer KGs could support this process by allowing users to select specific nodes and explore the graph accordingly. Indications of associated granularity tree or specific frames of reference FDOs aid in filtering the displayed triples or mind-map representations of statement FDOs, streamlining the user's exploration journey (Fig. 11).

These exploratory tools and visualizations can be combined in a seamless and intuitive manner, enabling users to switch between text-based, form-based, and graphical mind-map presentations of the data, providing users with a holistic and versatile approach to engaging with the information contained within a FAIRer KG (Fig. 11).

# Conclusion

What makes data and metadata FAIR and actionable for humans and machines alike? In this paper, I argue that two prerequisites must be distinguished to be able to answer the question of what makes data and metadata FAIR and actionable for humans and machines alike:

1. **Structural FAIRness of data and metadata**: There are specific structural prerequisites that are required for data and metadata to be FAIR and actionable, which include
    a. **technical properties**: Data and metadata must be structured in specific ways to be FAIR and actionable, including their format, underlying schema, and the vocabulary used, and these structural elements must be combined into an ideally broadly accepted standard. The **EOSC IF** addresses this aspect of FAIRness.
    b. **cognitive properties**: Data and metadata must be structured in such a way that humans can easily comprehend the meaning they carry. Adding **cognitive interoperability** as a fifth interoperability layer to the EOSC IF would address this aspect of FAIRness.

2. **Operational FAIRness of data and metadata**: Considering the large amounts of data that users often face when tackling a research question, users need tooling to support their data exploration, searching, and processing tasks. These **task-specific tools** must
    a. **support different exploration strategies**: **visual information seeking strategies** such as *'Overview first'*, *'Search first'*, and *'Details first'*, which also require some structural properties of data and metadata.
    b. **be usable**: Users must be able to find the tools and use them properly without too much effort involved. This aspect of FAIRness is also addressed by the **cognitive interoperability** of data and metadata.

With its focus on cognitive interoperability, we can infer from **operational FAIRness** additional structural requirements for FAIR data and metadata, leading to the proposal of the **Principle of human Explorability** of data and metadata as an additional principle for FAIR, resulting in the **FAIRer Guiding Principles**.

The overall goal of adding cognitive interoperability to the EOSC IF and proposing the FAIRer Guiding Principles is to address the key challenge of striking a balance between machine-actionability and human-actionability in information technology systems to enable efficient data management and exploration. I argue that reaching this goal requires understanding FAIRness not solely as a structural property of data and metadata, but also as depending on a comprehensive **FAIR Services ecosystem**. This ecosystem must include (55):

1. a **terminology service**, comprising a registry and look up service for terms defined in controlled vocabularies and for entity mappings across vocabularies for establishing terminological interoperability,
2. a **schema service**, comprising a registry and look up service for data schemata and schema crosswalks for establishing schema interoperability,
3. an **operations service**, comprising a registry and look up service for operations as readily executable code, each of which indicating the schemata used for their input and output data, for establishing readily executable **machine-actionability**.

Cognitive interoperability and human explorability of data and metadata are aspects of data management that attracted little attention in the context of KGs. With semantic units organized as FDOs, I introduce a concept that provides a framework that supports cognitive interoperability and that meets the specifications of the Principle of human Explorability. Semantic units add resources to FAIRer KGs that are of a higher level of abstraction than the low-level abstraction of resources used in the triples of conventional KGs. Technically, semantic unit resources and their corresponding FDOs instantiate corresponding FDO classes, while semantically, they represent statements or interconnected sets of semantically and ontologically related statements that are represented as semantically meaningful units of information. This organizational structure not only partitions the data graph layer of a FAIRer KG into different levels of representational granularity, granularity perspectives, and frames of reference, but also introduces a comprehensive framework for making statements about statements.

Furthermore, within FAIRer KGs, semantic units play a pivotal role in enhancing the FAIRness of data and metadata, because every semantically meaningful proposition is organized as a separate statement FDO that instantiates a corresponding statement FDO class. By referencing a data schema and corresponding query patterns in the statement FDO class for create, read, update, and delete (CRUD) tasks, each statement FDO references via its class affiliation the underlying data schema. Because every triple of the data graph layer of a FAIRer KG belongs to exactly one statement FDO, **all data in FAIRer KGs have information about their underlying data schema and how to query them**, thus contributing to the overall FAIRness of their data and metadata. This inherent structure not only aids users in finding specific data points of interest but also empowers developers with CRUD query patterns, all while **circumventing the need for developers and users to learn graph query languages** (*Challenge 2*).

Our proposed UI framework builds upon semantic units, offering novel ways of navigating and exploring FAIRer KGs, filtering for information of interest, and providing multiple entry-points for accessing and exploring the graph, zooming in and out across different levels of representational granularity, highlighting and making accessible any granularity trees and frames of reference in the data. Semantic units also contribute to **resolving the conflict and dilemma between machine-actionability and human-actionability of data and metadata** in a FAIRer KG (*Challenge 1*), since data representations can provide the contextual information required for machines, while humans can access and explore a graph of reduced complexity by utilizing graph-based displays based on dynamic mind-map patterns or form-based textual displays based on **dynamic labels**. Consequently, semantic units provide the framework required for UIs to align with the **visual information seeking strategies of *'Overview first'*, *'Search first'*, and *'Details first'***, providing additional structure that supports the development of new workflows and accompanying UIs for KG exploration approaches such as KG profiling and summarization, KG exploratory search, and KG exploratory analytics (*Challenge 3*).

By organizing statement units as **nanopublications**, forming **statement FDOs**, and by organizing compound units as **RO-Crates**, forming **compound FDOs**, semantic units with their taxonomy of different types of semantic units provide a promising framework for implementing different types of FDOs, lending itself to federated compound FDOs where the actual data in the form of statement FDOs can be distributed across different stakeholders, **possibly supporting explorability of data and metadata over a federated KG**.

In summary, semantic units lay the foundation for cognitive interoperability in FAIRer KGs, substantially increasing their terminological and propositional interoperability. With their ability to decouple data display from storage, they provide a powerful solution for presenting complex data to human users while ensuring machine-actionability. This innovation fosters a collaborative environment for sharing data among stakeholders, while data stewardship remains with domain experts or institutions (following Barend Mons' *data visiting* as opposed to *data sharing* (91)). The modular and nested nature of FDOs based on semantic units promotes broader accessibility and adoption for both software developers and users who lack expertise in semantics, thus fostering the wider FAIRification of data and metadata.

As we move forward, we envision a landscape where FAIRer KGs serve as dynamic and intuitive knowledge repositories, seamlessly supporting both human exploration and machine interaction, ultimately advancing the field of data management and knowledge representation.

# Abbreviations

| | |
|---|---|
| BFO | Basic Formal Ontology |
| EOSC | European Open Science Cloud |
| EOSC IF | European Open Science Cloud Interoperability Framework |
| FAIR | Findable, Accessible, Interoperable, and Reusable |
| FAIREr | Findable, Accessible, Interoperable, Reusable, and Explorability |
| FAIREr KG | FAIREr Knowledge Graph |
| FAIR KG | FAIR Knowledge Graph |
| FDO | FAIR Digital Object |
| FMA | Foundational Model of Anatomy Ontology |
| GDPR | General Data Protection Regulation |
| GO | Gene Ontology |
| GUPRI | Globally Unique Persistent and Resolvable Identifier |
| IAO | Information Artifact Ontology |
| ID | Identifier |
| IDO | Infectious Disease Ontology |
| KG | Knowledge Graph |
| NCI | National Cancer Institute |
| NCIT | NCI Thesaurus OBO Edition |
| NoSQL | Not only SQL |
| OBI | Ontology for Biomedical Investigations |
| OBO | Open Biological and Biomedical Ontology |
| OMIT | Ontology for MIRNA target |
| OWL | Web Ontology Language |
| PATO | Phenotype and Trait Ontology |
| RDF | Resource Description Framework |
| RO | OBO Relations Ontology |
| RO-Crates | Research Object Crate |
| SEMUNIT | Semantic Unit Ontology (currently in development) |
| SHACL | Shape Constraint Language |
| SPARQL | SPARQL Protocol and RDF Query Language |
| SQL | Structured Query Language |
| STATO | the statistical methods ontology |
| UBERON | Uber-anatomy ontology |
| UI | User Interface |
| UO | Units of Measurement Ontology |

# Declarations

## Ethics approval and consent to participate

Not applicable

## Consent for publication

Not applicable

## Availability of data and materials

Not applicable

## Competing interests

The authors declare that they have no competing interests

## Funding

Lars Vogt received funding by the ERC H2020 Project 'ScienceGraph' (819536) and by the project DIGIT RUBBER funded by the German Federal Ministry of Education and Research (BMBF), Grant no. 13XP5126B (TIB).

## Author's contributions

# Acknowledgements

I thank Philip Strömert, Roman Baum, Björn Quast, and Peter Grobe for discussing some of the presented ideas. I am solely responsible for all the arguments and statements in this paper. This work was supported by the Project 'DIGIT RUBBER', funded by the Federal Ministry of Education and Research of Germany (FZK: 13XP5126B) and the ERC H2020 Project 'ScienceGraph' (819536).